%% file: main.tex
\documentclass[12pt,final,fleqn]{article}

\usepackage[margin=1in] { geometry }
\usepackage{amssymb,amsmath, bm}
\usepackage{verbatim}
\usepackage[utf8]{inputenc}
\usepackage{setspace}
\usepackage{enumitem}
\usepackage[bottom]{footmisc}
\usepackage{url}
\usepackage[font={bf}]{caption}
\usepackage{float}
\usepackage{setspace}
\usepackage{latexsym}
\usepackage{graphicx}
\usepackage{marvosym}
\usepackage{amsmath} 
\usepackage{authblk}
\usepackage{xcolor}
\usepackage{blindtext}
\usepackage{changepage}
\usepackage[section]{placeins}

\usepackage[natbibapa]{apacite}
\bibpunct{(}{)}{;}{a}{}{,}

\usepackage{tabularray}
\UseTblrLibrary{booktabs}
\usepackage{longtable}
\usepackage{array}
\usepackage{multirow}
\usepackage{wrapfig}
\usepackage{float}
\usepackage{colortbl}
\usepackage{pdflscape}
\usepackage{tabu}
\usepackage{threeparttable}
\usepackage{threeparttablex}
\usepackage[normalem]{ulem}
\usepackage{makecell}
\usepackage{xcolor}
\usepackage{siunitx}
\usepackage{adjustbox}

\newcolumntype{d}{S[
    input-open-uncertainty=,
    input-close-uncertainty=,
    parse-numbers = false,
    table-align-text-pre=false,
    table-align-text-post=false
 ]}

\usepackage{tabularray}
\usepackage{float}
\usepackage{graphicx}
\usepackage[normalem]{ulem}
\UseTblrLibrary{siunitx}

\NewTableCommand{\tinytableDefineColor}[3]{\definecolor{#1}{#2}{#3}}

\usepackage{dcolumn}
\newcolumntype{.}{D{.}{.}{-1}}
\newcolumntype{d}[1]{D{.}{.}{#1}}
\usepackage{multirow}
\usepackage[figuresright]{rotating}
\usepackage{pdflscape}
\usepackage{subcaption}

\usepackage{CJKutf8}

\usepackage{grffile}
\usepackage{afterpage}
\usepackage{float}
\usepackage[section]{placeins}

\usepackage{theorem}
\theoremstyle{plain}
\theoremheaderfont{\scshape}

\newtheorem{hyp}{Hypothesis}

\usepackage{pdflscape}

\usepackage[compact]{titlesec}

\usepackage{color}
\usepackage{hyperref}
\usepackage{xcolor}
\hypersetup{
    colorlinks,
    linkcolor={blue!50!black},
    citecolor={blue!50!black},
    urlcolor={blue!80!black}
}

\newcommand{\aref}[1]{\hyperref[#1]{Appendix~\ref{#1}}}

\usepackage{minitoc}
\usepackage[toc,page,header]{appendix}

\usepackage{chngcntr}
\usepackage{etoolbox}
\usepackage{lipsum}



\allowdisplaybreaks[4]
\titleformat{\subsection}
  {\itshape\large}{\thesubsection}{1em}{}

\begin{document}
\singlespace
\title{\textbf{Foreign influencer operations: How TikTok shapes American perceptions of China \thanks{We thank Tony Cai for building the TikTok app and Jason Zhao for excellent research assistance. We also thank participants at conferences and workshops at Copenhagen Business School, the University of Pennsylvania, and the EPSA annual conference for valuable feedback. We are grateful for funding from the Amsterdam Institute for Social Science Research and The Whitney and Betty MacMillan Center for international and Area Studies at Yale. This project was approved by ethics advisory boards at the University of Amsterdam (FMG-12978) and Yale University (2000038305).}}\vspace{-1ex}}
\author{\vspace{0.75cm}Trevor Incerti\thanks{Assistant Professor, Department of Political Science, University of Amsterdam.} \vspace{-1ex}, Jonathan Elkobi\thanks{PhD Student, Department of Political Science, Yale University.} \vspace{-1ex}, and Daniel Mattingly\thanks{Associate Professor, Department of Political Science, Yale University.}\vspace{-1ex}}
\date{\vspace{0.75cm} \today}
\doparttoc % Tell to minitoc to generate a toc for the parts
\faketableofcontents % Run a fake tableofcontents command for the partocs
 
%\part{} % Start the document part
%\parttoc % Insert the document TOC
\maketitle
\pagenumbering{gobble}

% ABSTRACT -------------------------------------------------------------

\begin{abstract}
\noindent How do authoritarian regimes strengthen global support for nondemocratic political systems? Roughly half of the users of the social media platform TikTok report getting news from social media influencers. Against this backdrop, authoritarian regimes have increasingly outsourced content creation to these influencers.  To gain understanding of the extent of this phenomenon and the persuasive capabilities of these influencers, we collect comprehensive data on pro-China influencers on TikTok. We show that pro-China influencers have more engagement than state media. We then create a realistic clone of the TikTok app, and conduct a randomized experiment in which over 8,500 Americans are recruited to use this app and view a random sample of actual TikTok content. We show that pro-China foreign influencers are strikingly effective at increasing favorability toward China, while traditional Chinese state media causes backlash. The findings highlight the importance of influencers in shaping global public opinion.
\end{abstract}
%These messages have traditionally been circulated abroad by foreign-facing state media outlets. 

%\vspace{0.5cm}
%\begin{center}
%\textcolor{red}{EARLY DRAFT: Please do not cite or circulate}
%\end{center}

\vspace{0.5cm}
\begin{center}
\end{center}

\pagebreak
%\doublespace
\pagenumbering{arabic}
\setcounter{page}{1}

%%%%%%%%%%%%%%%%%%%%%%%%%%%%%%%%%%%%%%%%%%%%%%%%%%%%%%%%%%%
% MAIN DOCUMENT
%%%%%%%%%%%%%%%%%%%%%%%%%%%%%%%%%%%%%%%%%%%%%%%%%%%%%%%%%%%

%\section*{Introduction}

\noindent Authoritarian regimes regularly attempt to shape global public opinion towards non-democratic political systems using foreign influence operations. Traditionally, these influence operations have been conducted through state media organs, among other channels, such as Russia Today or CGTN, which is the news channel of the state-run China Global Television Network \citep{mattingly2025chinese}. However, regimes also now outsource content creation to ``influencers'' who are not explicitly tied to the regime and produce clickbait content that seeks to entertain as well as persuade \citep{lu2022pervasive, mattingly2022soft}. These foreign influencers---who create and disseminate entertaining social media content highlighting positive aspects of a particular country---have been discussed by state media \citep{chen2025, xinhua2025, seitz2022china} and major journalistic outlets \citep{brumfiel2025, bloomberg2025}, identified as a threat by  governments \citep{charon2021chinese, ryan2023singing}, and even fostered diplomatic disputes \citep{koh2025}.

%We posit that foreign influencers represent a new form of influence operations, and define foreign influencers as individuals who create and disseminate entertaining social media content highlighting the positive cultural, economic, or political aspects of a particular country, and/or the negative cultural, economic, or political aspects of a rival country. Whether this content is government-sponsored ``propaganda'' is unclear. Some influencers have direct financial and other ties to the regime; others have no ties whatsoever, but produce content that is useful for the regime or matches the goals of official state media organs.

Understanding political influencers is important because increasingly large shares of Americans get political information from social media creators instead of traditional news outlets. As of 2025, 37\% of Americans used TikTok, making it America's third most used social media platform.\footnote{Excluding YouTube.} 63\% of Americans age 19-28 use TikTok, and roughly half of this age group use TikTok daily \citep{gottfried2025americans}. In addition, roughly half of American TikTok users---or 17\% of U.S. adults---report regularly getting news on TikTok. Only 1\% of accounts followed by Americans are institutional news outlets, and TikTok users are as likely to report getting news from influencers as news outlets \citep{wang2024closer}.  

Despite the increasing attention paid to foreign influencers on video platforms like TikTok, little research to date has examined their prevalence or effectiveness. Previous research has established the persuasive ability of Chinese state-owned media abroad \citep{mattingly2025chinese}. Other work has shown that the Chinese government has used influencers in a strategy of ``decentralized propaganda'' aimed at domestic audiences \citep{lu2025decentralized, lu2022pervasive}, and has examined Russian influence operations on Twitter \citep{golovchenko2020cross, eady2023exposure}. One recent study on TikTok focuses on domestic influencers and shows how they shape attitudes towards American politics, but it remains unclear if the power of influencers extends to international politics \citep{ChmelEtAl2024}.

Importantly, influencers who disseminate pro-state narratives for their own profit offer a low-cost alternative to traditional foreign influence operations.  If creation of this content is both economically viable for creators and achieves the same goals of disseminating persuasive pro-state narratives as traditional state media, it could represent a new and minimal cost tool in states' attempts to influence global public opinion. However, it remains unclear whether: (1) foreign influencer content reaches a wide audience, (2) the creation of foreign influencer content is a lucrative endeavor for content creators, and (3) foreign influencer content is as persuasive as traditional state media.

%However, it is far from clear if influencers supporting authoritarian governments on video platforms like TikTok are effective. It is possible that these messengers may have lower engagement than traditional state media, which is supported with large budgets and may have slicker production values. It is also possible that these messengers may be ineffective outside of the specific context of, for example, Chinese domestic politics. 

In this paper, we first collect descriptive data on viral pro-China content appearing on TikTok and find that influencers are more prevalent than state media in users' feeds. We show that videos created by foreign influencers outperform those created by state media across virtually all engagement metrics. Engagement with this content is significant --  roughly 10\% of influencer videos have reached 2.5 million views. However, compared with the larger body of non-political content on TikTok, even pro-China influencers appear in user feeds relatively rarely.  

%We identify the most prevalent creators of this content and analyze the main themes used by pro-China influencers. Overall, we find that influencers are just as likely to focus on political messages as state media (albeit with differences in the specific content). 

%We estimate that the median pro-China content creator makes roughly $400-$800 USD per month from TikTok, suggesting that producing pro-China content can be a way to earn modest amounts of money. In short, foreign influencers are popular relative to state media but not popular enough to support large numbers of lucrative accounts. 

Next, we examine whether the content created by pro-China influencers on TikTok shapes Americans' attitudes towards China by building our own version of the TikTok app and conducting a randomized experiment with over 8,500 participants. The app we created and controlled had many of the same qualities as TikTok, most notably the ability to scroll seamlessly through hundreds of videos. Its general appearance and interface also mimicked the TikTok app. In our experiment, we randomly inserted pro-China foreign influencer content into some feeds and Chinese state media into others, while a third group only received a placebo set of non-political videos.\footnote{Note that before joining the study, all participants were informed that they might be exposed to content from the Chinese government, and then given an option to consent or not. After the study, they were also debriefed about the content they watched.}  We find that users were less likely to scroll away from foreign influencer content relative to some the most popular celebrity accounts on TikTok. We also find that exposure to foreign influencer videos  increased favorability towards China and increased positive views of China's culture, economy, and political system. By comparison, watching state media decreased favorability towards China. We conclude that the effectiveness of influencers is likely a combination of the messenger and the message --- the most effective influencer videos generally focused on positive messages about how advanced the Chinese economy is. 

Our findings show how authoritarian regimes, and China in particular, have begun to export the home-grown strategy of using influencers to shape public opinion, with potential global consequences for democracy \citep{hyde2020democracy}. We build on several related bodies of social science research. Previous research suggests that content designed to be entertaining --- from reality TV shows to influencer accounts on social media --- can be especially persuasive \citep{kim2023entertaining, kim2025mirage, kim2025american,  kluser2025entertainment}. Other work shows how authoritarian regimes increasingly use a strategy of outsourcing content creation to influencers targeted at domestic audiences \citep{lu2021capturing, lu2022pervasive, lu2025decentralized, mattingly2022soft, liang2024useful}. We show how authoritarian regimes have now begun to export this strategy --- leading to a growing number of ``foreign influencer operations.''

%This suggests that entertaining content created by influencers may naturally reach a wider audience while still containing persuasive narratives. Recent research also shows that those \textit{least} likely to choose to watch content are most persuaded by it, suggesting that algorithmic promotion of content to new viewers could both broaden reach and cause attitude change \citep{egamiplacebo}. 

%However, influencer content may also be less effective at shifting opinions in a pro-state direction than official state media, as it is typically less explicitly tied to state narratives. In short, foreign influencers may be less effective at achieving the state's goals on a per video basis, but reach a wider audience, generate more engagement, and come at virtually no cost to the regime. 

%We find no differences between videos produced by Chinese or non-Chinese influencers.

%Next, we show that foreign influencers are significantly more effective at increasing favorability toward China than state media, with state media even causing backlash and reducing favorability of China among respondents. This effect holds when users are able to scroll through videos

%We test two channels for how influencer propaganda shapes public opinion: source credibility and emotional resonance. 
\section{Theories of Influence Operations on Social Media}

Authoritarian states spend large sums on foreign influence operations intended to shape global public opinion, but the consequences of these influence operations remains a matter of debate.\footnote{For example, China earmarked \$6.6 billion USD for overseas state media in 2009. See among others: \cite{colley2023news, dukalskis2021making, hartig2020rethinking, moore2024two}.}  On the one hand, one study found that exposure to a Russian influence operation on Twitter in the 2016 election had no discernible effects on attitudes or behavior \citep{eady2023exposure}. Another study found relatively limited effects of the YouTube algorithm on policy attitudes \citep{liu2025short}. This might lead one to expect small or null effects for influence operations on video platforms like TikTok.  On the other hand, other studies have found that viewing video content from outlets like China's CGTN can shift attitudes towards China by large amounts \citep{mattingly2025chinese, egamiplacebo}. Influencers on TikTok have also been shown to have substantial effects on attitudes towards American politics in a randomized field experiment \citep{ChmelEtAl2024}. This might lead us to expect positive and meaningfully sized effects for the pro-China influencers we study.

Building on the second set of studies, we expected that both Chinese state media and pro-China influencers would be likely to have substantial positive effects on pro-China attitudes. This led to our first pre-registered hypothesis:
\begin{hyp}
Viewing videos produced by pro-China influencers and state media will increase respondent affinity for China relative to the placebo condition. 
\label{hyp: affinity_placebo}
\end{hyp}

There are also reasons to expect that influencers might be more effective than traditional media at changing attitudes, beliefs, or behavior. Audiences on platforms like TikTok are turning towards influencers as important sources of news, suggesting they are more engaging and entertaining but not necessarily less effective at changing minds \citep{ChmelEtAl2024}. Indeed, entertainment media more broadly has been shown to have durable effects on attitudes and beliefs \citep{kim2025american, kim2025american, kim2023entertaining}. Authoritarian regimes have also turned to producing ``clickbait'' content and ``soft propaganda'' to influence domestic audiences, suggesting that it is an effective tool of persuasion \citep{lu2021capturing, mattingly2022soft}. This leads to our second pre-registered hypothesis:
\begin{hyp}
Viewing videos produced by pro-China influencers will cause larger shifts in affinity for China than viewing videos produced by Chinese state media.
\label{hyp: affinity_state}
\end{hyp}

Moreover, the battle for attention on video platforms is fierce, and so we might expect that influencers will be especially effective at capturing and holding audience attention relative to state media. Prior studies have shown that Chinese state media is effective at changing minds, but that few people chose to watch it, and those who do watch it already have strong positive affinity towards China \cite{egamiplacebo}. A response to this problem---that state media is effective when watched but has low natural engagement---is to turn to influencers who are not saddled with the baggage of being labeled state media, and whom may be able to craft more naturally compelling messages. In China, the regime's messaging strategy appears to acknowledge this. On Douyin, the Chinese version of TikTok, the central party apparatus has turned to a strategy of ``decentralized propaganda,'' in which it farms out content creation to local governments and local party organizations to allow for the most engaging propaganda to compete and rise to the top of the algorithm \citep{lu2025decentralized}. This leads to our third major pre-registered hypothesis:

\begin{hyp}
In the free-choice arm [where respondents are free to scroll between videos], compliance for those assigned to pro-China influencers will be higher than compliance for those assigned to watch state media.
\label{hyp: compliance}
\end{hyp}

In our results, we found support for the first and second hypotheses. Influencers increased respondent affinity towards China. However, we found an unexpected negative effect for state media on affinity towards China. We did not find support for the third hypothesis---respondents watched influencers and state media at roughly equal rates, albeit with opposite effects on favorability towards China. However, we did find that users engaged with pro-China influencer videos longer than the placebo videos drawn from the top (mostly American) accounts on TikTok.

\section{The pro-China foreign influencer landscape on TikTok}

\subsection{Data collection}

To study how TikTok might influence attitudes towards China, we first collect descriptive data on the most prevalent pro-China content being produced on the TikTok platform. We collect engagement data (likes/diggs, shares, comments, views,  descriptions, and hashtags) for all videos that appear in a TikTok search for ``\#China'' for the month of February 2025. We collect this data using 10 new accounts that are created from scratch and launched from major cities in the US, using new cookies prior to search execution in order to reduce the risk of TikTok's algorithms oversampling pro or anti-China content in our searches after learning our preferences. We then employ snowball sampling using the open followers' accounts to find more influencers with content regarding China. This returns a total of approximately 40,000 videos representing about 23,000 unique accounts.

Next, for each unique user, we examine the first two pages of videos of their content. If at least 15 percent of these videos contained the term ``China,'' we manually examine each account. We categorize the user as a foreign influencer if at least 3 of the 10 most recent videos were about China, they were not known Chinese state media accounts, \textit{and} they positively portrayed Chinese politics, economics, and/or culture in their videos. In addition, we examine which accounts these users follow, and if these accounts meet the same criteria as above we also categorize them as foreign influencers. This ultimately leads to a total of 204 of the original list of over 20,000 accounts that use ``\#China'' in any video being classified as ``foreign influencers'' according to the criteria above. Next, among the foreign influencer accounts, we manually classify the accounts as covering one or more of the following three topics: Chinese politics, Chinese economics, or Chinese culture.\footnote{Culture includes topics such as travel, food, or humor based content.}\footnote{Given the video-based nature of the content, text-based topic modeling is not possible. While videos contain text summaries, they are often not entirely indicative of the video content.}  Finally, we extract all 3321 videos and associated metadata from these foreign influencer accounts and 316 videos and metadata from official state media accounts.

\vspace{0.25cm}
\input{tables/descriptive_table}

\subsection{Findings}

\begin{figure}[H]
\includegraphics[width = \textwidth]{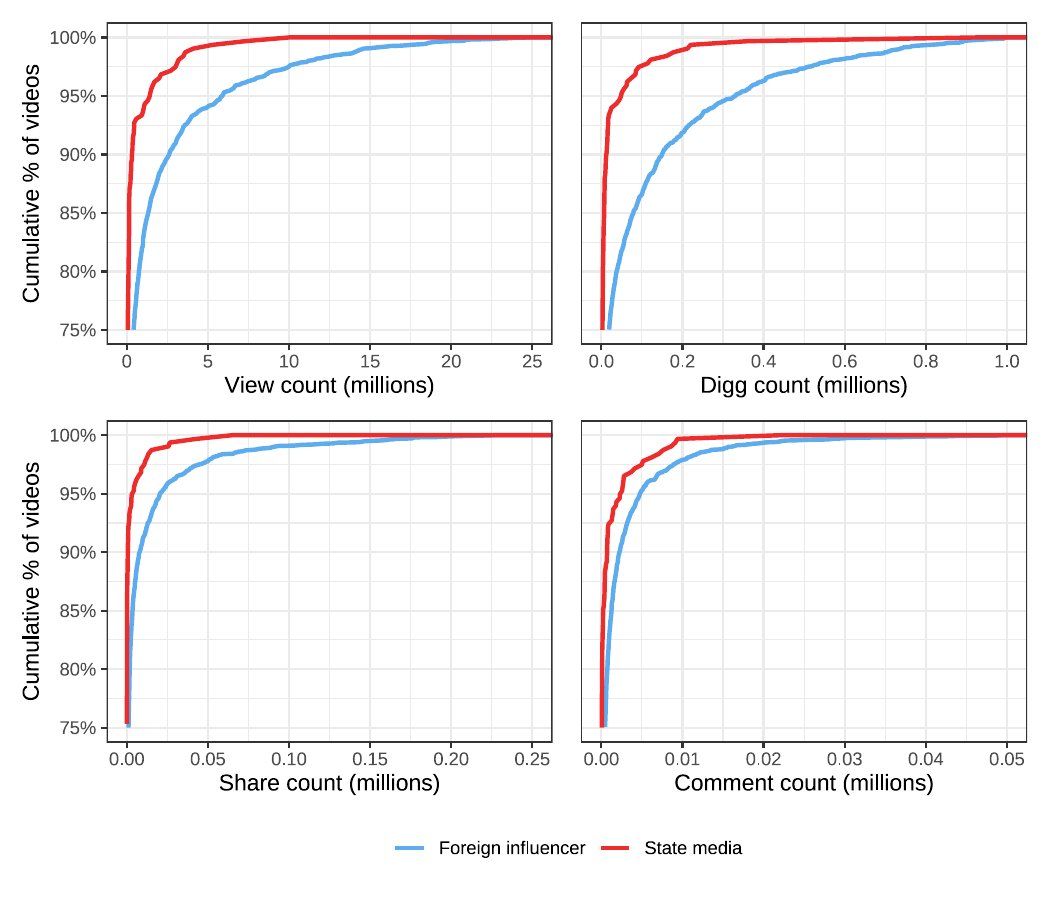}
\caption{Empirical cumulative distribution functions of video engagement metrics for foreign influencers and state media}
\small
\vspace{-0.3cm}
\label{fig: ecdf}
\end{figure}

First, we investigate how foreign influencer accounts compare to state media accounts in terms of overall levels of engagement. Specifically, we examine total numbers of views, diggs (i.e., ``likes''), shares, and comments for all videos created by influencers and state media. \autoref{fig: ecdf} presents an empirical cumulative distribution function of video views, and shows what percentage of videos created by influencers and state media have more than N number of total views. \autoref{fig: ecdf} demonstrates that videos created by foreign influencers outperform those created by state media across all engagement metrics. For example, roughly 10\% of influencer videos have reached 2.5 million views or more, while only 5\% of state media videos have reached similar view counts. Foreign influencer accounts also average higher engagement than state media accounts overall (\autoref{fig: ecdf}). This confirms that viewers are more likely to consume and interact with content produced by influencers than by official state-run media accounts across a sample of accounts observed in the real-world. Later we will also explore which content individuals choose to consume when experimentally presented with different content types. Finally, we estimate that foreign influencers generate median monetary returns of between \$400-\$800 USD per month from creating foreign influencer content (\autoref{fig: earnings}).\footnote{See \nameref{sec: monitary_returns} in the appendix for additional details.}

%Next, we estimate the thematic content of all foreign influencer and state media videos. We classify videos based on whether they are on the topic of culture, economics, or politics using the \textit{activeText} algorithm developed by \citet{bosley2025improving}, which estimates the probability documents belong to a specific class following manual labeling and model training by researchers, and focuses manual labeling on documents with the most uncertainty.\footnote{Note: we classify videos using the text descriptions of videos provided by the video authors.} We find that foreign influencers are slightly more likely to create videos about politics than state media (\autoref{fig: fi_sm_topics}). % We therefore confirm that foreign influencers posses much higher natural engagement than state media, but are less likely to talk explicitly about political and economic topics than official state media. 

%\begin{figure}[H]
%\includegraphics[width = \textwidth]{figures/fi_sm_topics.pdf}
%\caption{Estimated percentage of topics discussed by foreign influencers and state media}
%\small
%\vspace{-0.3cm}
%\label{fig: fi_sm_topics}
%\end{figure}

\section{The effect of foreign influencers on attitudes about China}

\subsection{Experimental design}

Next, we estimate the effect of pro-China influencer content on attitudes and behavior using an original experimental design that mirrors the TikTok platform (an overview of the experimental design can be found in \autoref{fig: experiment_overview}). Our design builds on \citet{mattingly2025chinese}, who survey audiences in 19 countries and show that watching CGTN shifts attitudes towards China, as well as \cite{egamiplacebo} who find that those who are most persuaded by CGTN are those who are least likely to choose to watch it.

\begin{figure}[H]
    \centering
        \begin{subfigure}[t]{0.45\textwidth}
        \centering
        \includegraphics[width=3cm]{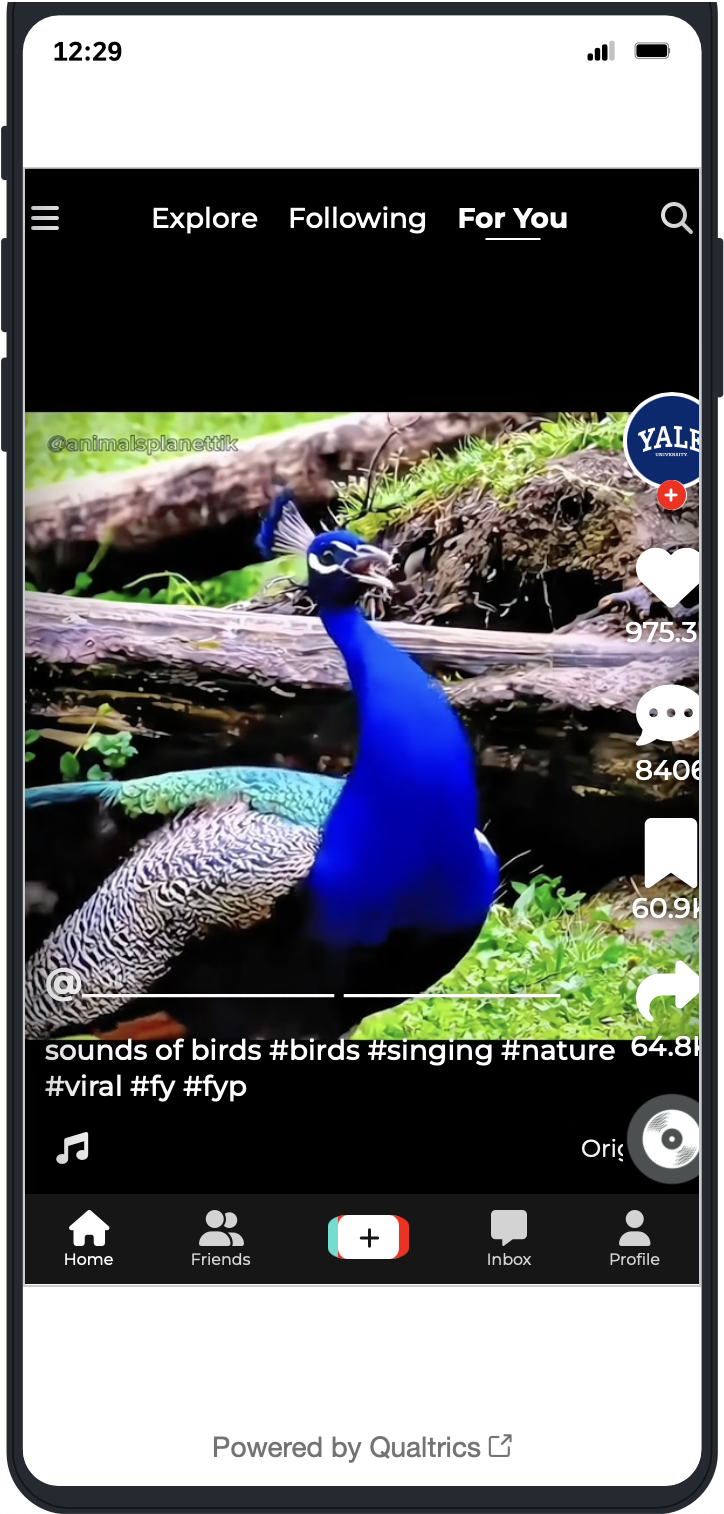}
        \caption{Placebo} \label{subfig: placebo}
    \end{subfigure}
    ~
    \begin{subfigure}[t]{0.45\textwidth}
        \centering
        \includegraphics[width=3cm]{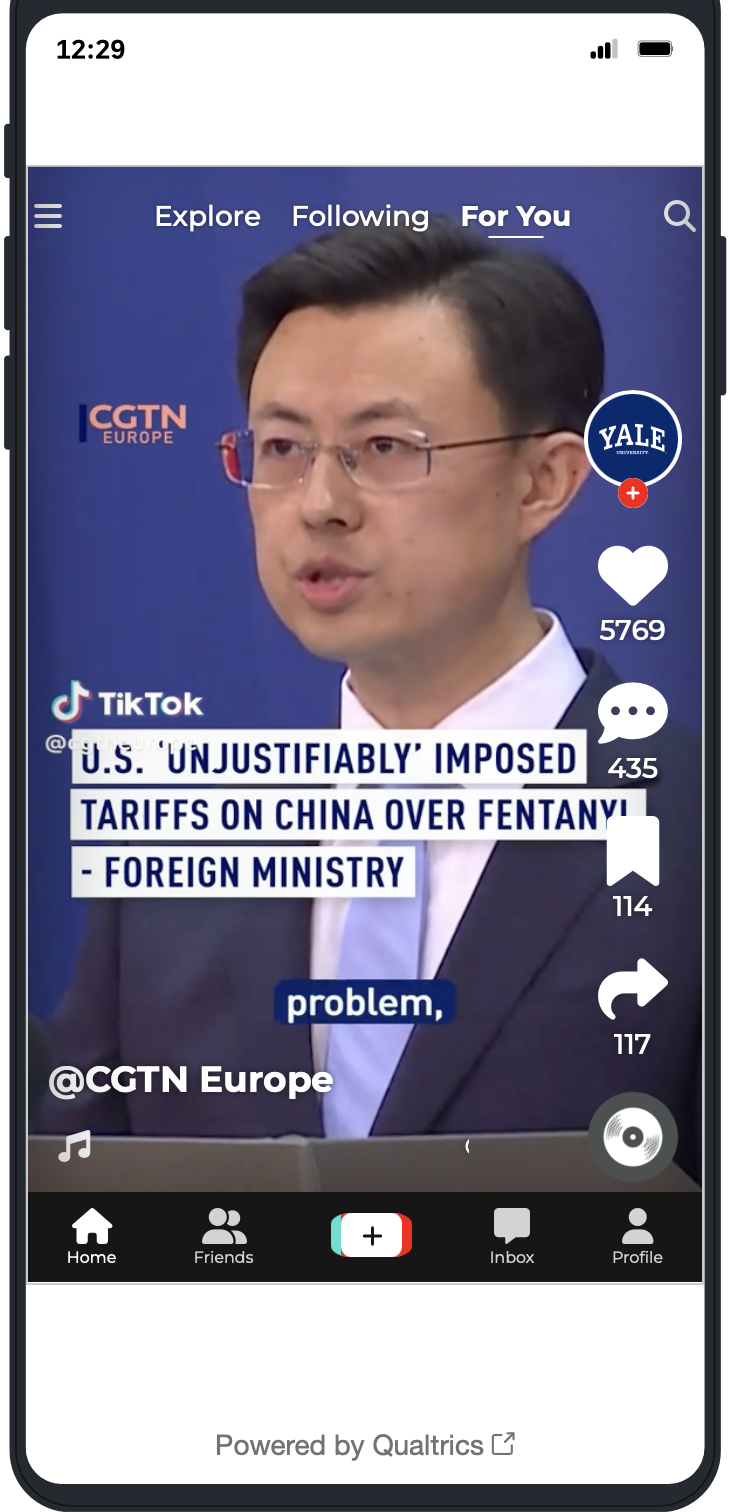}
        \caption{State media} \label{subfig: state_media}
        %\vspace{0.5cm}
    \end{subfigure}
    ~ 
    \begin{subfigure}[b]{0.3\textwidth}
        \centering
        \includegraphics[width=3cm]{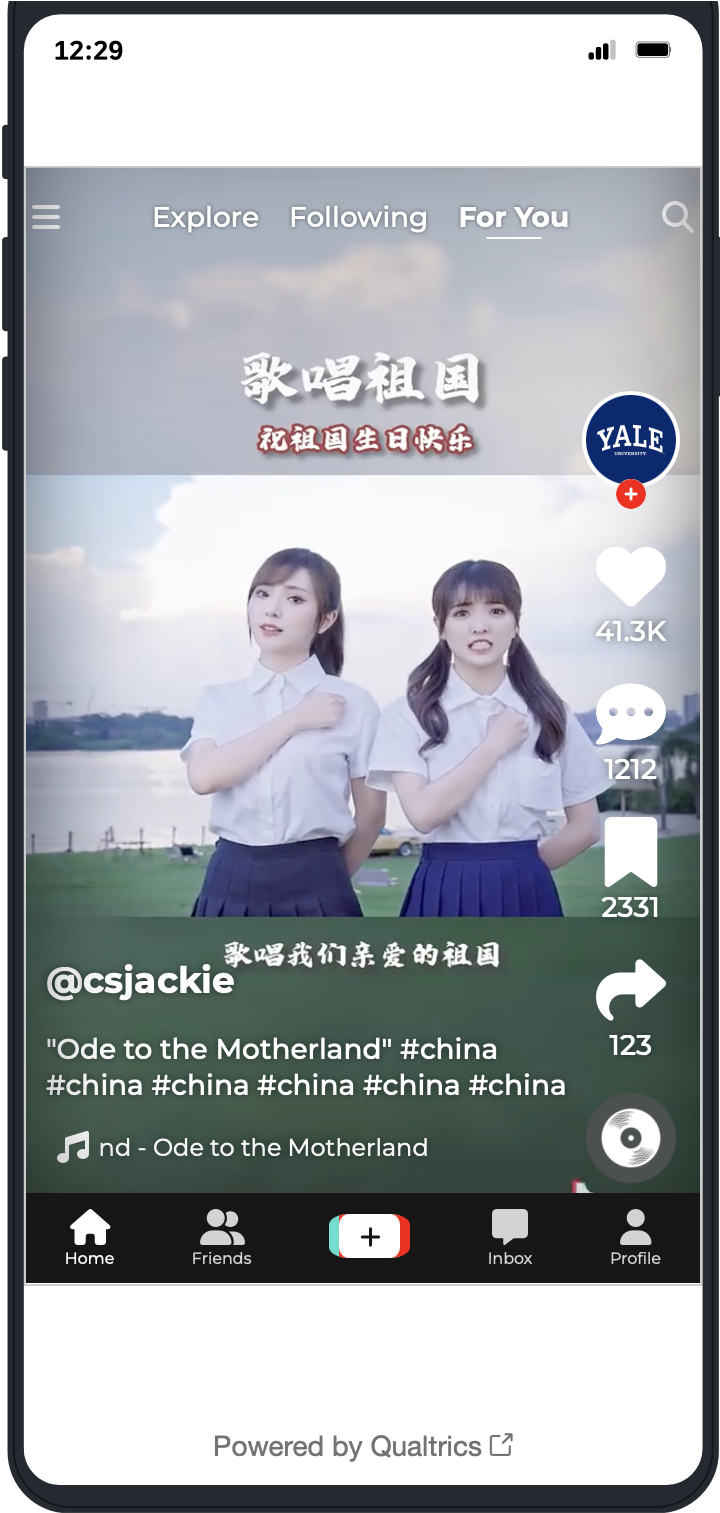}
        \caption{Culture} \label{subfig: culture}
    \end{subfigure}
    ~
    \begin{subfigure}[b]{0.3\textwidth}
        \centering
        \includegraphics[width=3cm]{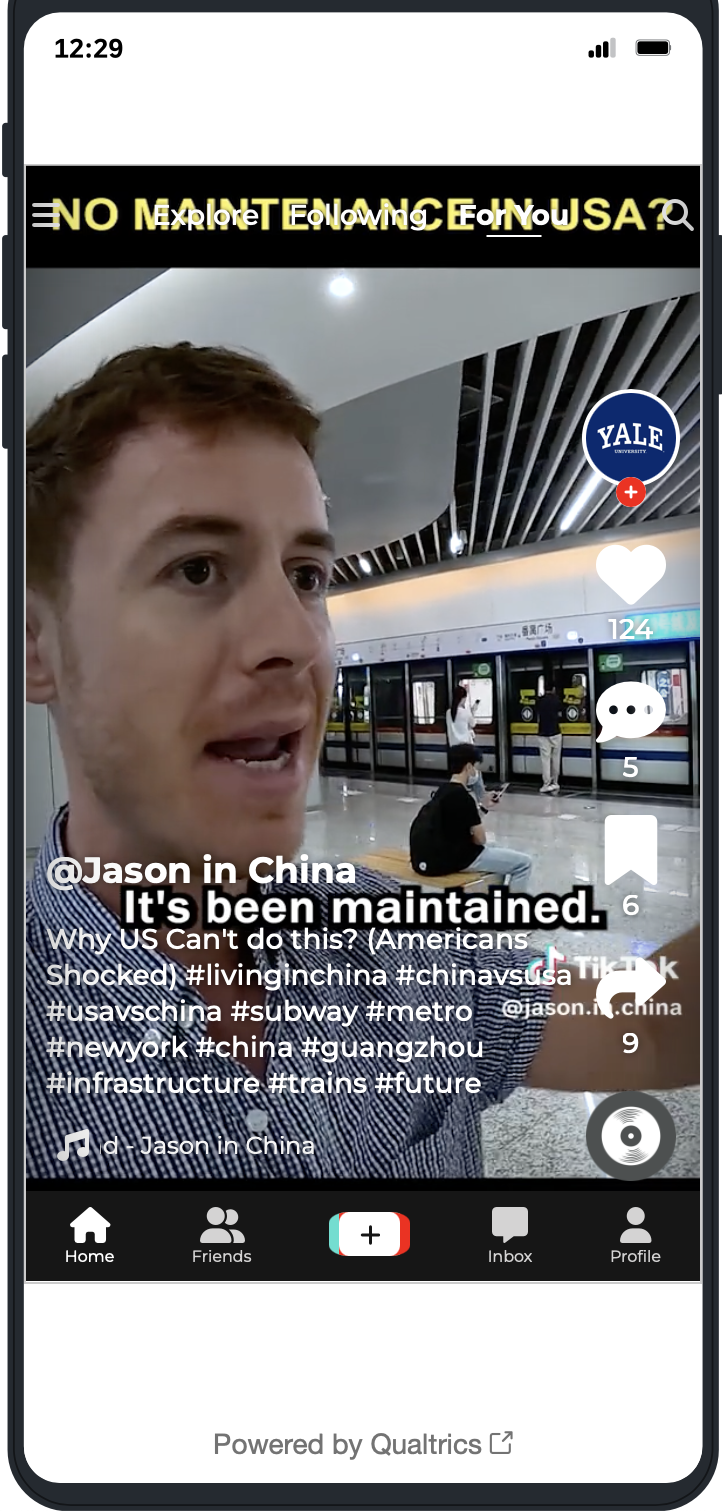}
        \caption{Economy} \label{subfig: economy}
    \end{subfigure}
    ~
    \begin{subfigure}[b]{0.3\textwidth}
        \centering
        \includegraphics[width=3cm]{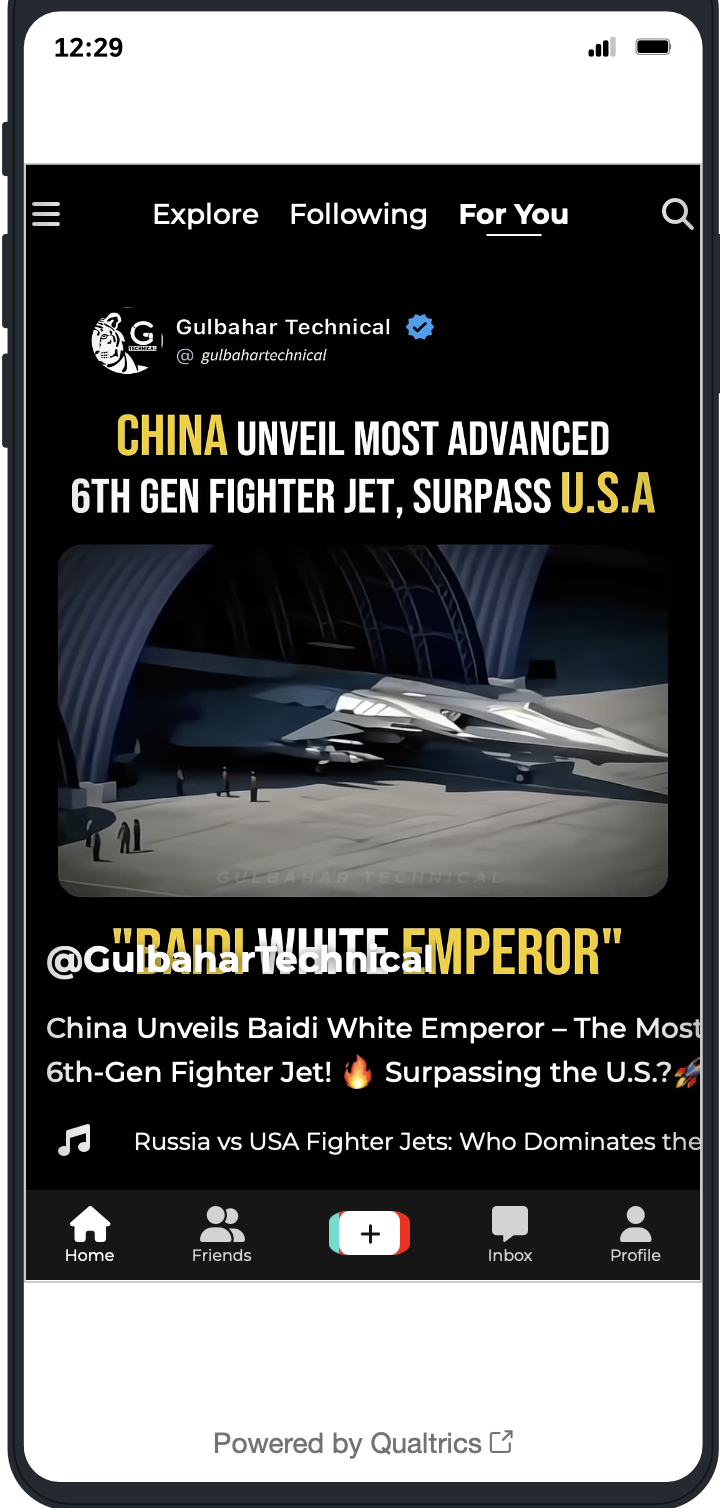}
        \caption{Politics} \label{subfig: politics}
    \end{subfigure}
    ~
    \caption{Example treatment videos}\label{fig: treatments}
\end{figure}

The experiment takes the following steps. First, we recruit a diverse sample of 8,600 US respondents balanced on age, gender, ethnicity, and region from Cint.\footnote{Formerly Lucid.} After obtaining informed consent, in which users are told they may be asked to watch content from the Chinese government, we expose participants to a baseline survey where we collect information about respondent characteristics. We then direct respondents to our clone of the TikTok app (see \autoref{fig: treatments} for an example). Next, we randomly assign participants to one of two conditions---\textit{Forced} or \textit{Free Choice}---with 20\% probability of assignment to the \textit{Forced} group and 80\%  probability of assignment to the \textit{Free Choice} group. %(25\% to the \textit{Forced} group and 75\% to the \textit{Free Choice} group).\footnote{We randomly assign more respondents to the \textit{Free Choice} group as we expect smaller treatment effects from this group, and therefore wish to increase sample size and therefore power in this group.}

Participants in the \textit{Forced} group are randomly assigned to a random sample of real TikTok videos from Chinese state media accounts,\footnote{These are: CGTN, CGTN America, CGTN Europe, Pheonix TV Hong Kong, the Chinese Embassy in the United States, and ``iamlijingjing.''} pro-China influencer accounts, or placebo videos unrelated to China or politics. Each respondent was asked to watch for 4 minutes. Within the state media and influencer groups, pro-China videos are randomly assigned with equal probability of assignment from a larger corpus of videos on politics, economics, and culture. Drawing on our observational analysis, we sampled 100 videos weighted by view count, and manually eliminated videos that  were in languages other than English, that were not about China, or that contained inappropriate language or imagery. Our final pool included 67 influencer videos and 89 state media videos.\footnote{The influencer feeds were more likely than state media to include some videos not related to China.} Respondents in the \textit{Forced} group \textit{must watch all of the videos} (i.e., there is no option to skip videos).

%Pro-China videos are block randomly assigned to ensure subjects receive an equal mix of videos on politics, economics, and culture.\footnote{Videos are randomly assigned from a corpus of [N] videos.}

Participants in the \textit{Free Choice} group are free to watch or skip any videos they choose---as they would on the real TikTok platform---again for a duration of 4 minutes. Respondents are then randomly assigned to either: (1) a random sample of entertainment-related placebo videos, (2) a random sample of placebo videos \textit{and} videos from  Chinese state media accounts, or (3) a random sample of placebo videos \textit{and} a random sample of videos from pro-China influencer accounts. Pro-China videos are again randomly assigned with equal probability from larger corpora of videos on politics, economics, and culture. Respondents in the \textit{Free Choice} group \textit{can freely choose to watch or skip any of the videos}, but are required to spend at least 4 minutes before proceeding. An overview of the experimental design can be found in \autoref{fig: experiment_overview}.

%As in the \textit{Forced} group, pro-China videos are block randomly assigned to ensure subjects receive an equal mix of videos on politics, economics, and culture.\footnote{Videos are randomly assigned from a corpus of [N] videos.}

Following treatment, respondents complete an endline survey with the following (pre-registered) primary outcome variables: favorability towards China, attitudes towards China's economy, and attitudes towards China's political system (see appendix for exact survey wording).\footnote{We also collect data on secondary outcome variables such as: whether respondents prefer the Chinese or American economic and/or political models. The full list can be found in the appendix.}  As a behavioral measure of willingness to take political action related to China, we ask respondents to sign a petition memorializing the Tiananmen incident and condemning the Chinese government.\footnote{We track whether respondents clicked a link to sign the petition to measure petition signing.} Finally, respondents are exposed to a debrief explaining the design and purpose of the experiment, as well as offering alternative viewpoints from those presented in the videos.\footnote{See \nameref{sec: debrief} in the appendix for a full transcript.} 

\subsection{Estimation procedures}

Our primary estimands are intent-to-treat (ITT) effects---the average treatment effects of random assignment to watch (1) pro-China influencers or (2) state media videos on the outcomes listed above when compared to the Placebo group. We estimate the ITT using ordinary least squares (OLS) with HC2 robust standard errors and include the following pretreatment covariates: \textit{gender}, \textit{age}, \textit{education}, \textit{national pride}, and \textit{left-right political orientation}.\footnote{For robustness, we also report randomization inference p-values and perform multiple comparisons corrections. We conduct multiple comparisons corrections using the Bonferroni, Holm, Hochberg, Hommel, Benjamini Hochberg, and Benjamini Yekutieli procedures.} We also estimate heterogeneous treatment effects (HTEs) for three pre-registered subgroups---age, education, and left-right political alignment---by regressing the outcome variables above on treatment assignment separately for each subgroup,\footnote{Also known as conditional average treatment effects (CATEs), or an ATE specific to a subgroup of subjects.} and hypothesize that attitudes will shift more for younger, less educated, and more extreme respondents.

%The ability to skip videos in the \textit{Free Choice} condition is equivalent to revealing compliance status. We therefore define compliers as any respondent who watches 1 or more pro-China videos,\footnote{Compliance can also be thought of as an increasing function of number of videos viewed. We therefore formalize this definition of compliance in the appendix, as well calculate effects as a function of level of compliance according to this definition.} and calculate both intent-to-treat (ITT) effects and complier average causal effects (CACE) for the \textit{Free Choice} group. A formal definition of compliance in the context of the experimental design can be found in \nameref{sec: compliance} in the appendix. % We therefore also calculate the complier average causal effect of watching a pro-China influencer video on the outcomes listed above when compared to the Placebo group, and the complier average causal effect of watching a pro-China state media video on the outcomes listed above when compared to the Placebo group. 

The ability to watch or skip pro-China videos in the free choice treatment arms is equivalent to compliance in a one-sided noncompliance framework. As subjects can choose to watch any duration of pro-China videos they choose, this represents a continuous measure of compliance sometimes referred to as variable treatment intensity. \citet{angrist1995two} show that two-stage least squares (2SLS) applied to a causal model with variable treatment intensity and nonignorable treatment assignment identifies a weighted average of per-unit treatment effects. We therefore estimate the weighted average local average treatment effect (LATE)---i.e., the marginal effect of an additional minute of watching pro-China videos---using an instrumental variables/2SLS estimation strategy in which we use treatment assignment as an instrument for time spent watching pro-China videos.

All procedures described in this section were preregistered.\footnote{The full preregistration document is provided in the appendix.} 

\subsection{Effect of foreign influencers on attitudes towards China}

\begin{figure}[H]
\centering
\begin{subfigure}[b]{\textwidth}
    \centering
    \includegraphics[width=\textwidth]{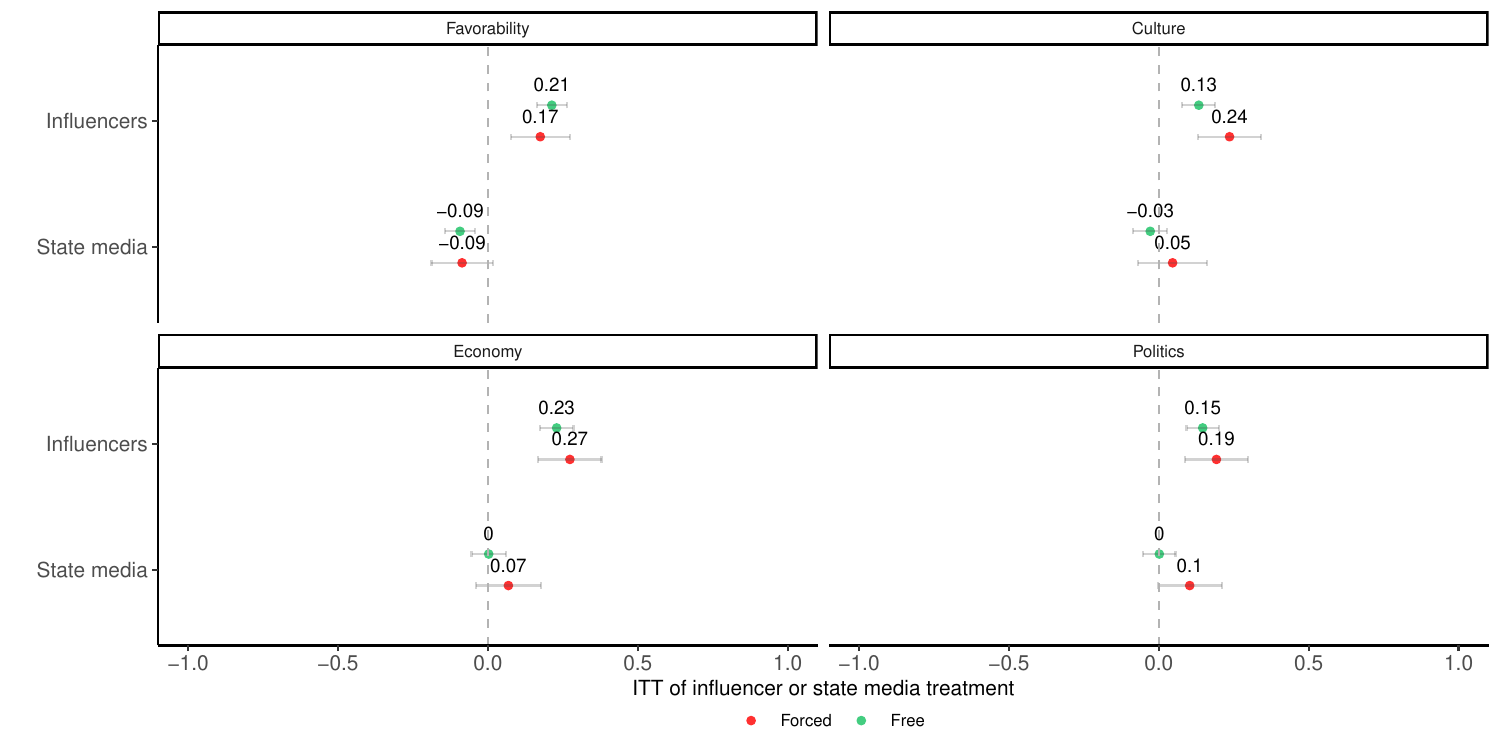}
    \caption{Intent-to-treat effects of influencer and state media videos on attitudes towards China} \label{subfig: combined_plot_itt}
\end{subfigure}
%\vspace{-0.5cm}
\begin{subfigure}[b]{\textwidth}
    \centering
    \vspace{-0.5cm}
    \includegraphics[width=\textwidth]{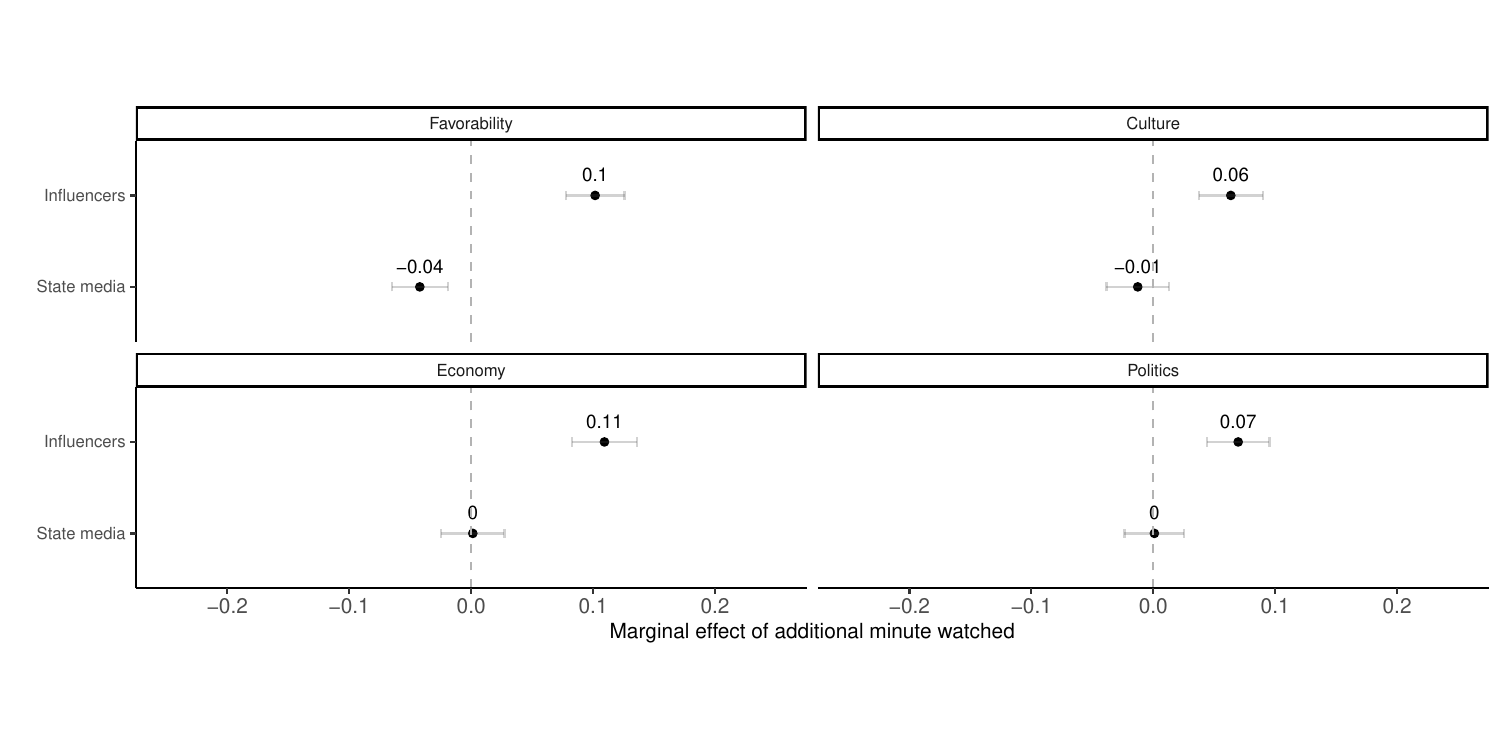}
    \vspace{-0.5cm}
    \caption{Marginal effect of additional minute of influencer and state media videos watched on attitudes towards China} \label{subfig: late}
\end{subfigure}
\caption{Effects of influencer and state media videos on attitudes towards China} \label{fig: combined_figures}
\end{figure}

\autoref{subfig: combined_plot_itt} shows that pro-China influencer content appears effective at increasing favorability towards China (\textit{Hypothesis \autoref{hyp: affinity_placebo}}). Assignment to both the free choice and forced foreign influencer groups increased favorability of China among respondents. Moreover, influencer content increased perceptions of the strengths of China's economy, political system, and culture, with influencers appearing particularly effective at convincing respondents that China has a strong and technologically advanced economy (\autoref{fig: all_secondary}).\footnote{These findings are robust to calculation of p values using randomization inference, as well as even the most conservative multiple comparisons corrections (\autoref{tab: ri} and \autoref{tab: mcc}).}

By contrast, state media on TikTok is ineffective at increasing favorability toward China (\textit{Hypothesis \autoref{hyp: affinity_state}}), and in fact may even lead to backlash. Assignment to the free choice state media group significantly reduced favorability, while a comparable in magnitude but nonsignificant\footnote{p-value = 0.14 based on OLS with covariate adjustment or 0.2 using randomization inference without covariate adjustment. Recall that only 20\% of respondents were randomly assigned to the \textit{Forced} group, reducing sample size relative to the \textit{Free Choice} group.} decline was observed in the forced state media group. Consistent with the idea that exposure to state media caused backlash, respondents randomly assigned to both the forced and free choice state media arms reported higher levels of anger compared to the placebo group.\footnote{Respondents assigned to the state media arms also reported being less ``inspired'' and ``excited.'' While these emotion outcomes were pre-registered as exploratory outcomes, we note that all of these effects remain significant at the 1\% level even with the most conservative multiple comparisons corrections applied.} State media had no effect on perceptions of China's economy, political system, or culture. The sole domain in which state media nudged perceptions in the Chinese state's desired direction was foreign policy attitudes, where both influencers \textit{and} state media increased respondent support for trade and security cooperation. Only influencers, however, reduced perceptions that China is an enemy of the US, and influencers were still more effective than state media at shifting foreign policy attitudes (\autoref{fig: foreign_policy}).\footnote{This finding should be interpreted with caution, however, as it does not survive all multiple comparisons corrections (\autoref{tab: mcc}).} 

No content had an impact on respondent likelihood of signing a petition condemning the Tiananmen incident (\autoref{fig: petition}), and we observe little heterogeneity in response to treatment by age, education, or political ideology (\autoref{fig: hte}).  

The more content is consumed the stronger are the effects on average (\autoref{subfig: late}). The marginal effect of watching an additional minute of state media content on China favorability is negative 0.05 on our 4-point scale, while the marginal effect of an additional minute of influencer content is positive 0.1. Viewing additional influencer content similarly increases positive perceptions of China's culture, economy, and political system, but viewing additional state media has virtually no effect, further confirming hypotheses \autoref{hyp: affinity_placebo} and \autoref{hyp: affinity_state}.

\subsection{What do individuals choose to watch?}
 
Over the course of the four minute experiment, over half of individuals in the free choice group consumed at least 50\% of 3 or more pro-China videos, and over half of individuals consumed at least 75\% of 2 or more videos (\autoref{fig: watch_rates_x}). \autoref{fig: violin} shows that when respondents are free to pick and choose videos at will---as with the real TikTok platform---they on average watch a greater percentage of pro-China influencer and state media content than entertainment related content. Pro-China influencer videos are watched to a mean of 28\% completion, while state media videos are watched to 27\% completion and placebo influencer videos to 20\% completion.\footnote{Nature videos are the most popular, being watched to a mean of 34\% completion.} Watch rates vary by age group, with individuals under 30 watching pro-China influencer videos to a mean of 19\% completion compared to 32\% for those over 30, and individuals under 30 watching state media to a mean of 18\% completion vs. 32\% for those over 30 (\autoref{fig: violin_dem}). These results are not consistent with our third pre-registered hypothesis\footnote{That ``compliance for those assigned to pro-China influencers will be higher than compliance for those assigned to watch state media.''}, but it is notable that both pro-China influencer and state media content were consumed significantly more than the most popular entertainment related content on TikTok, in line with previous research that content about political outgroups drives engagement \citep{rathje2021out}.

\subsection{Which content is most persuasive?}

 We next estimate the individual effect of each video on participants’ favorability toward China. We randomly assign participants not only to treatment arms but also to specific videos. However, there is an important caveat: because of potential spillover effects between videos in each treatment arm, the estimates for individual videos cannot be interpreted naively as the causal effect of the individual video. While this limits our ability to estimate an unbiased causal effect of each video relative to placebo, it does help us understand the relative effect of each video within each treatment group. \autoref{tab:positive_videos} and \autoref{tab:negative_videos} summarize the top 15 most influential positive and negative videos for the free choice arm. 

The influencer videos with large positive effects generally highlight China's economic performance and dynamism. These videos focus on China's modern cities, electric vehicles, and infrastructure. We find no significant difference in treatment effects between Chinese influencers and American or European influencers, contrary to previous studies that have found co-nationals more persuasive \citep{liang2024useful}. Before correcting for multiple comparisons, there were many videos with significant positive effects, but after applying the most conservative correction in our pre-registration plan---Bonferroni---the top 13 videos have significant estimates.

As the video descriptions suggest, the negative effects of the state media arms are driven by videos that either criticize the United States or showcase Chinese power. For example, two of the videos with the largest negative effects depict the Chinese Ministry of Foreign Affairs (MOFA) criticizing America's domestic and foreign policies (e.g., tariffs, protests, and immigration). A second set of videos showcases China's military, politics, and drones.\footnote{Applying the Bonferroni correction, only the top 4 videos remain significant at the 5\% level. However, this is likely an overly conservative correction as the Bonferroni correction assumes complete independence and the most negative videos are topically similar.} 

Overall, these results suggest that differing content and differing messengers may both play a role in the persuasiveness of influencers relative to state media. The influencer content focuses largely on positive messages whereas the state media messages also include criticisms of the United States and more focus on China's growing military power. It may also be that influencers are also more credible messengers than state media for the same content since they are not explicitly aligned with the state. In reality, these are bundled treatments. Our study emphasizes realism at the cost of being able to pull apart these factors by, for example, creating artificial treatments that manipulate one dimension of difference. This is a potential avenue for future research.

\section{Discussion and conclusion}

Prior research shows that authoritarian regimes invest heavily in traditional media to shape global public opinion, and that under some conditions these channels can shift opinion \citep{mattingly2025chinese, egamiplacebo, dukalskis2021making}. Yet these outlets often face public skepticism and low engagement. By contrast, our results show that pro-China influencers on TikTok enjoy high levels of engagement and, in a naturalistic experiment using a simulated version of the TikTok app, we find that (1) this engagement emerges naturally even without algorithmic amplification, and (2) foreign influencers have substantial effects on attitudes towards China.

Our evidence builds on research that shows the persuasive power of entertainment media and on influencers in democratic contexts \citep{ChmelEtAl2024, kim2025mirage}. It also extends work that shows how the Chinese Communist Party uses soft propaganda and influencers to mold public opinion domestically \citep{lu2025decentralized}. Our results show how this strategy of outsourcing propaganda work to influencers has been exported to a global audience.

We remain cautious about the scope of these findings. Our experiment evaluates short-term exposure.  The extent to which the results we show persist remains unknown, and there may be general equilibrium effects, in which more wide-scale exposure to these messages could reduce their effectiveness. Moreover, we do not find any effects on our chosen behavioral measure, signing a petition. Existing studies of algorithmic recommendation systems and social media show that exposure to media produces limited polarization and opinion change \citep{liu2025short, eady2023exposure}.

However, modest shifts in attitudes delivered through repeated exposure can be consequential, especially on platforms like TikTok, which prioritizes engagement. TikTok influencers are also an increasingly important source of news for Americans. Continuing to study influence operations on these platforms is therefore important. Traditional monitoring efforts that target official foreign outlets will miss messages carried by actors with no formal affiliation with authoritarian states but overlapping incentives.

Future work can examine how these dynamics evolve over time, how they differ by platform, and how they persist. Over time the strategy of foreign influencers on TikTok may change as they learn about user preferences. At the same time, users may adapt in response to what they learn from foreign influencers. It is possible that the sale of the TikTok could lead to changes in the algorithm that may alter engagement with pro-China videos. Finally, it remains an open question how these attitudes persist and whether repeated exposure over a period of days and months changes the dynamics we observe here. Nevertheless, foreign influencers will most likely continue to be an important tool in the arsenal of non-democratic governments as they seek to mold global public opinion.

%\citep{wang2024closer}

%We postulate that Chinese state media on TikTok may be ineffective for four primary reasons. First, state media videos are labeled as such. Second, state media often contains news clips that are not necessarily pro-China in nature. Third, while state media increasingly attempts to emulate influencer content, it may be less effective at organically creating such content. Fourth, \citet{mattingly2025chinese} find that Chinese state media was least persuasive among North American\footnote{Not including the US.} and European respondents---Americans could be similarly unresponsive to state appeals.

%Our findings suggest that the effects of exposure are continuous and compounding, and even small increments in viewing time can lead to meaningful shifts in attitudes toward China. Just one minute of exposure to influencer videos increased participants’ favorability toward China by more than 1.4 standard deviations above the baseline mean. Conversely, additional exposure to state media content appears to exert a modest negative influence on favorability. Although our experiment does not measure the persistence of these effects over the long term, the short term change shows the importance of content type and engagement in shaping perceptions. Future research can investigate the persistence of these favorability shifts and, in addition, whether repeated or sustained exposure bolsters the short-term effects.

%%%%%%%%%%%%%%%%%%%%%%%%%%%%%%%%%%%%%%%%%%%%%%%%%%%%%%%%%%%
% REFERENCES
%%%%%%%%%%%%%%%%%%%%%%%%%%%%%%%%%%%%%%%%%%%%%%%%%%%%%%%%%%%
%TC:ignore
\pagebreak
\pdfbookmark[1]{References}{References}
\bibliography{bibliography}

%%%%%%%%%%%%%%%%%%%%%%%%%%%%%%%%%%%%%%%%%%%%%%%%%%%%%%%%%%%
% APPENDIX
%%%%%%%%%%%%%%%%%%%%%%%%%%%%%%%%%%%%%%%%%%%%%%%%%%%%%%%%%%%

\singlespace
\newpage
\appendix
\addcontentsline{toc}{section}{Appendix} % Add the appendix text to the document TOC
\part{\large{Supporting Information} }% Start the appendix part

\vspace{0.5cm}
\begin{center}
\large{\textbf{Foreign influencer operations: How TikTok shapes American perceptions of China}} \\ 
\end{center}
\vspace{0.25cm}
\parttoc 

\setcounter{table}{0}
\renewcommand{\thetable}{A\arabic{table}}
\setcounter{figure}{0}
\renewcommand{\thefigure}{A\arabic{figure}}
\pagenumbering{arabic}% resets `page` counter to 1
\renewcommand*{\thepage}{A\arabic{page}}

\normalsize
\newpage

\section{Supplemental material}

\subsection{Additional descriptive statistics}

\begin{figure}[H]
\includegraphics[width = \textwidth]{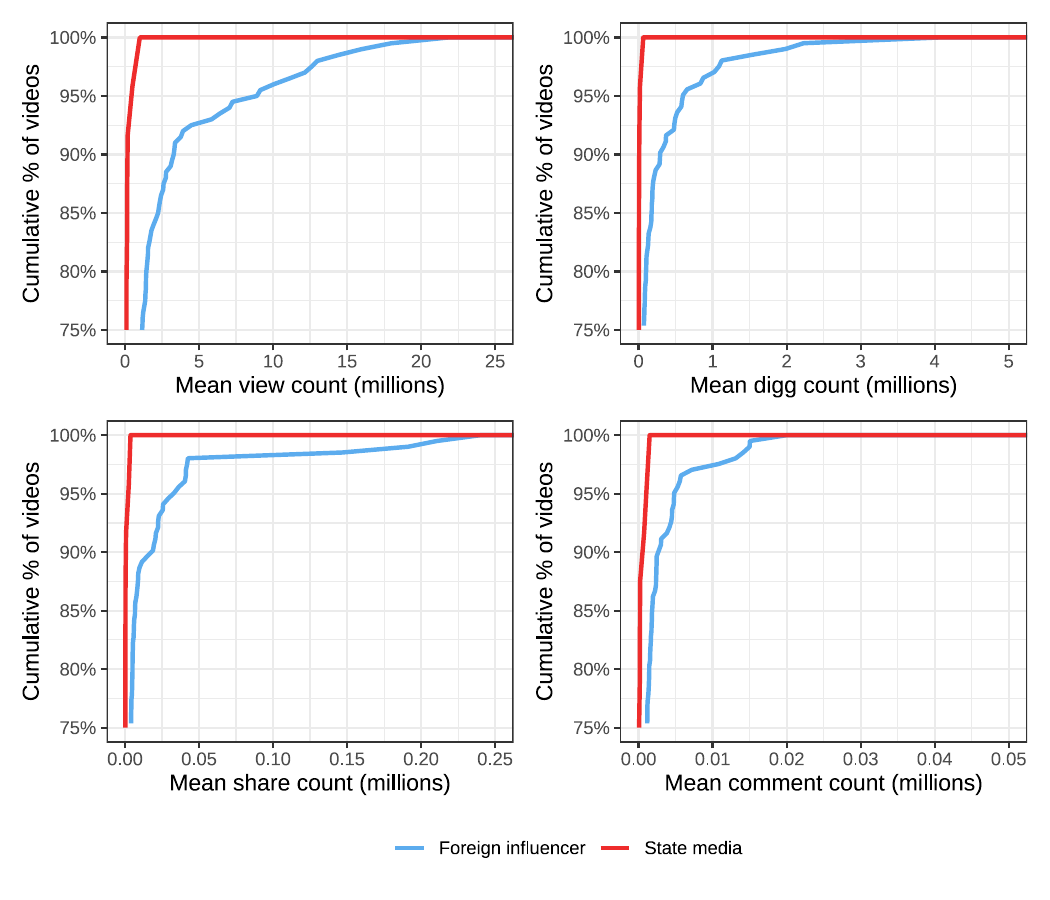}
\caption{Empirical cumulative distribution functions of average engagement metrics per author for foreign influencers and state media accounts}
\small
\vspace{-0.3cm}
\label{figures/fig: author_ecdf}
\end{figure}

\subsection{Experimental design}

\begin{figure}[H]
\includegraphics[width = \textwidth]{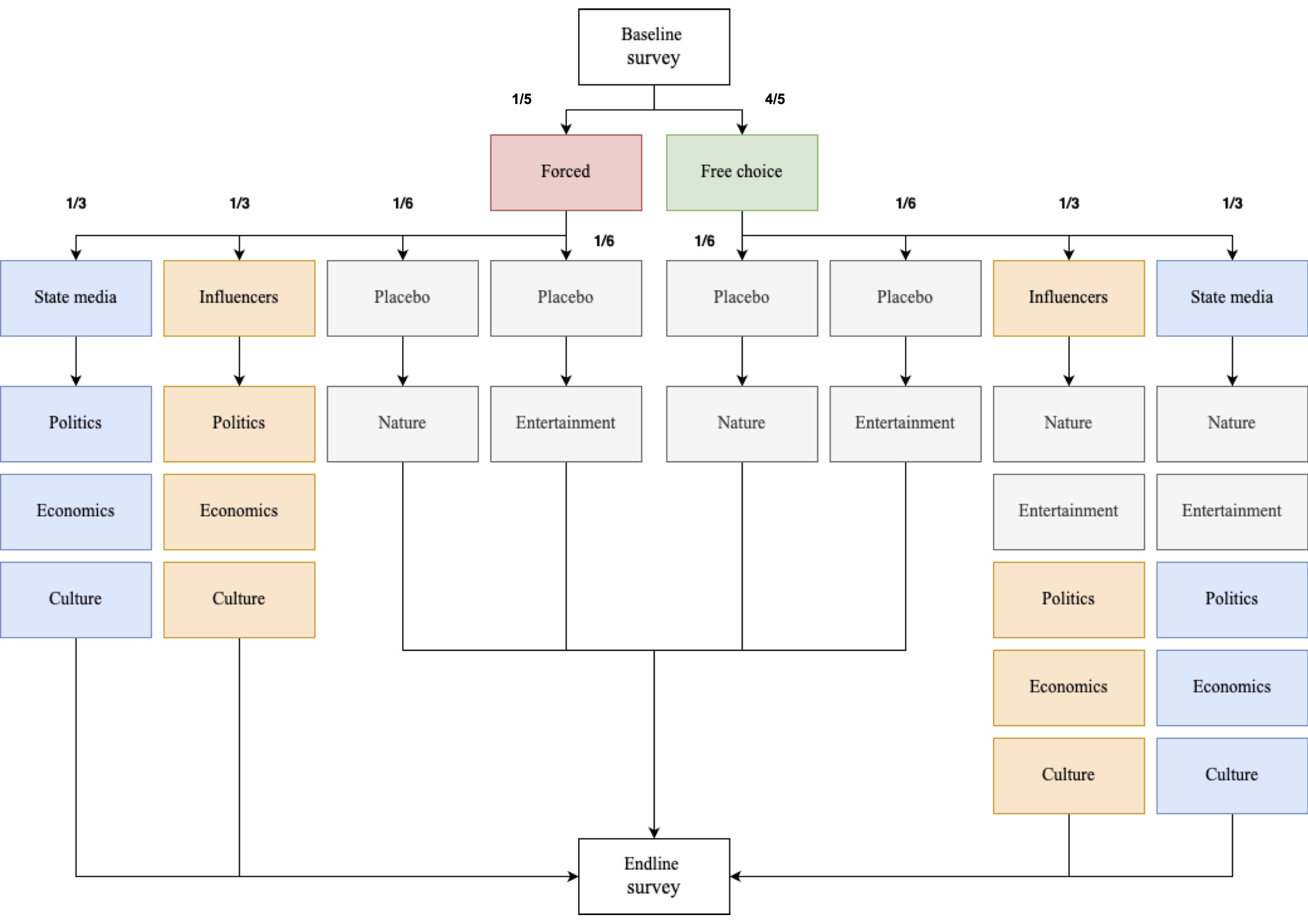}
\caption{Overview of experimental design. }
\small
\vspace{-0.3cm}
\label{fig: experiment_overview}
\singlespace
\end{figure}

\clearpage
\subsection{Data collection and respondent sample notes}

\subsubsection{Pilot study}

We began data collection with an initial pilot study of 500 respondents in July 2025. Results from the pilot sample can be found below, and were used to refine the final pre-registration document. Changes between the pilot and full sample pre-registration are as follows:

\subsubsection
{Data collection issues}

We note here minor technical issues that arose during response collection that a affected a small number of respondents. Some respondents were assigned to two treatment groups (i.e., they watched two sets of videos) due to internet connection issues and/or refreshing the experiment midway through the study. This affected a total of 8 respondents, who were subsequently removed from the analysis sample.

\clearpage
\subsection{Additional experimental figures}

\subsubsection{Descriptive statistics}

\begin{figure}[H]
\centering
\includegraphics[width = \textwidth]{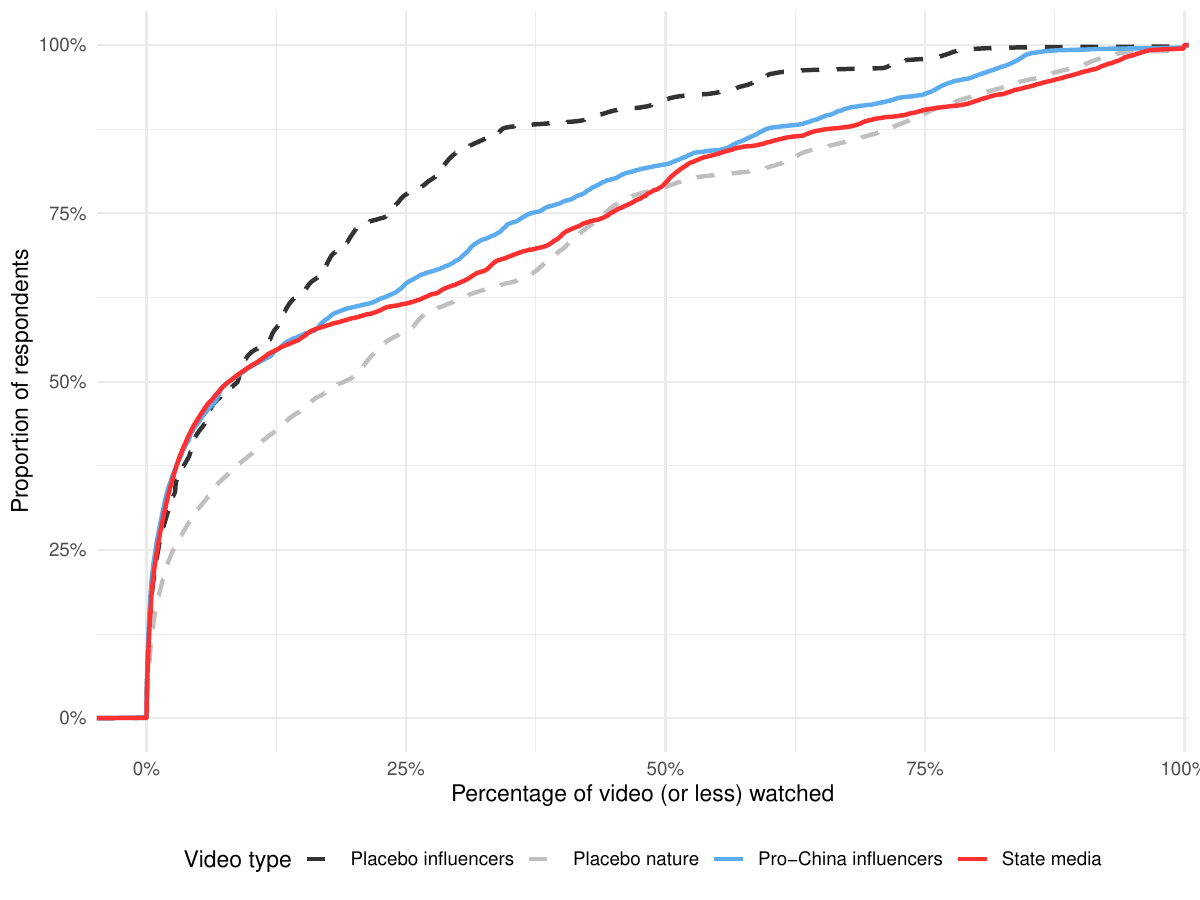}
\caption{Empirical cumulative distribution functions of percentage of videos (or less) watched by respondents, by video type}
\small
\vspace{-0.3cm}
\label{fig: video_ecdf}
\end{figure}

\begin{figure}[H]
\centering
\includegraphics[width = \textwidth]{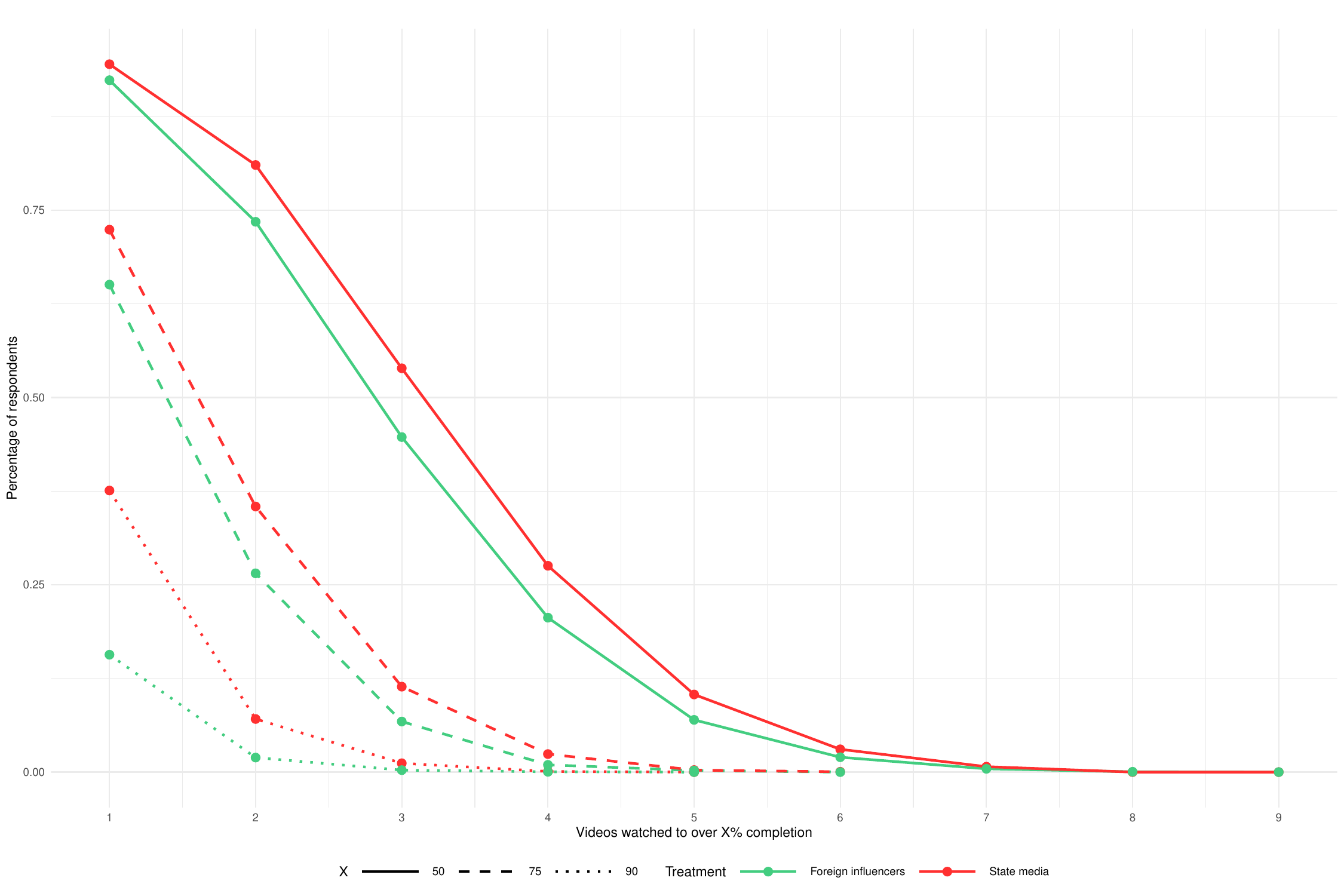}
\caption{Percentage of respondents who watched N videos to over X\% completion, by treatment group}
\small
\vspace{-0.3cm}
\label{fig: watch_rates_x}
\end{figure}

\begin{figure}[H]
\centering
\includegraphics[width = \textwidth]{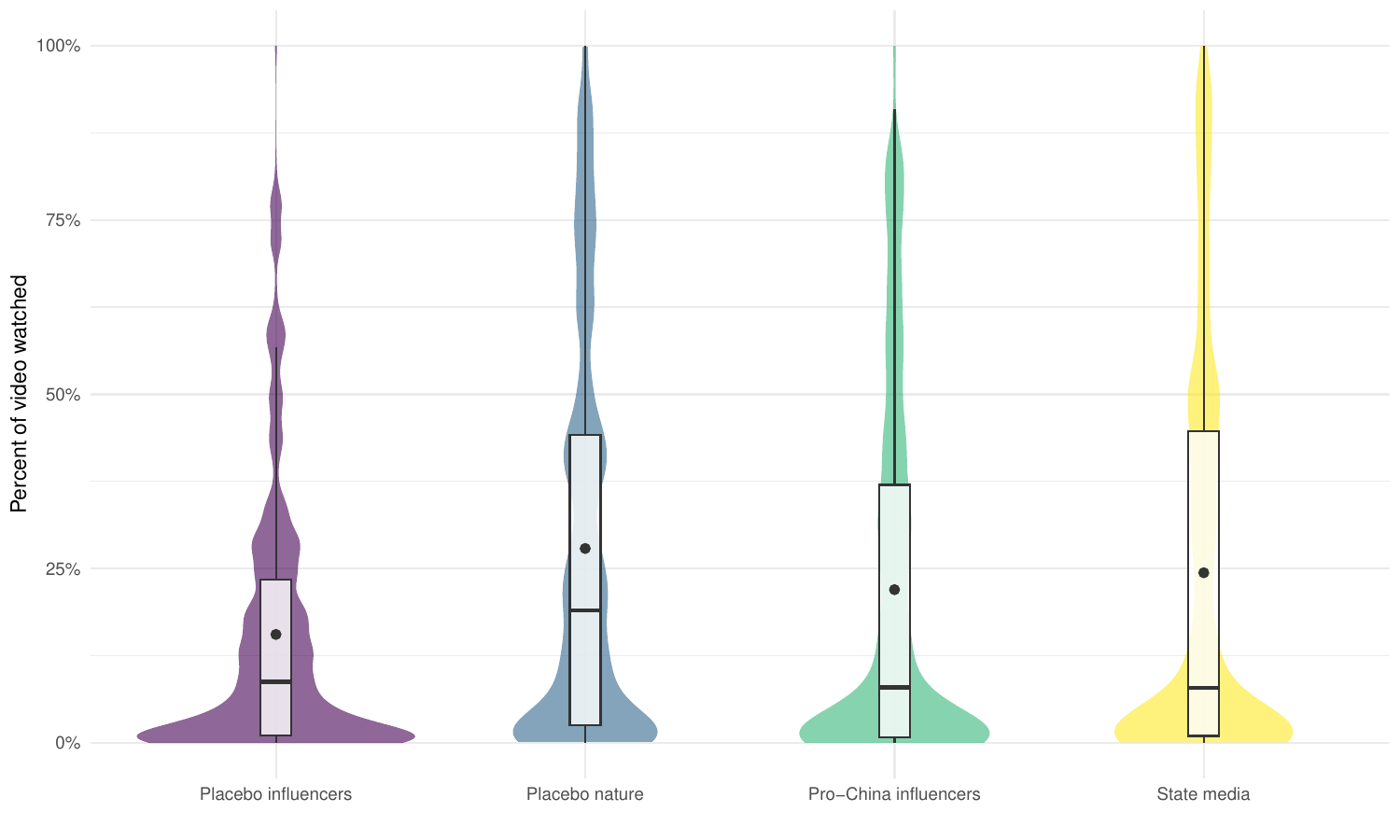}
\caption{Distribution of percentage of videos watched by treatment group}
\small
\vspace{-0.3cm}
\label{fig: violin}
\singlespace
\noindent
\footnotesize
\textit{Note}: Black line represents median and grey diamond represents mean.
\end{figure}

%\begin{figure}[H]
%\centering
%\includegraphics[width = \textwidth]{figures/video_violin_dem.pdf}
%\caption{Distribution of percentage of videos watched by treatment group}
%\small
%\vspace{-0.3cm}
%\label{fig: violin_dem}
%\singlespace
%\noindent
%\footnotesize
%\textit{Note}: Black line represents median and grey diamond represents mean.
%\end{figure}

\begin{figure}[H]
\centering
\includegraphics[width = \textwidth]{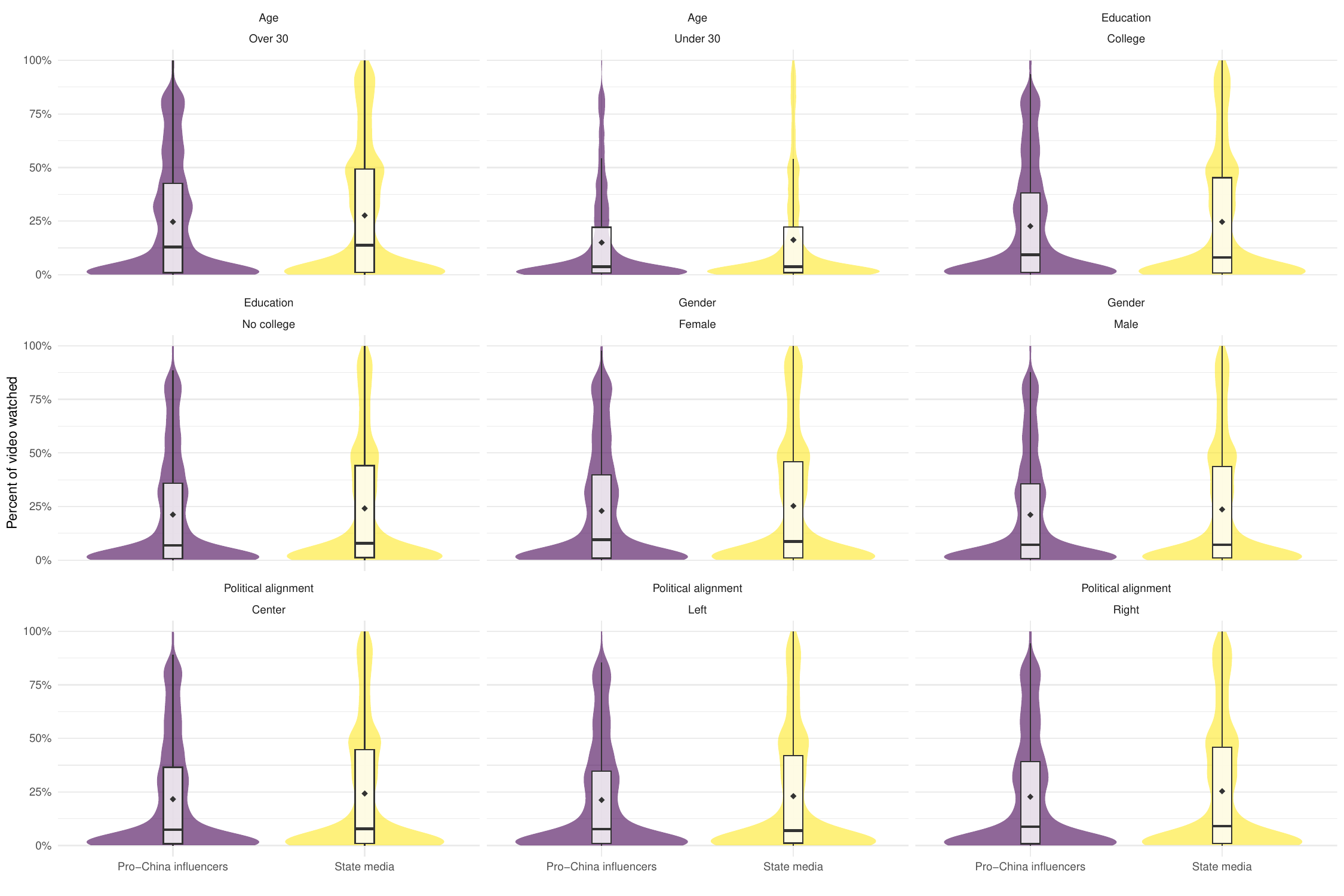}
\caption{Distribution of percentage of videos watched by treatment group}
\small
\vspace{-0.3cm}
\label{fig: violin_dem}
\singlespace
\noindent
\footnotesize
\textit{Note}: Black line represents median and grey diamond represents mean.
\end{figure}

\clearpage
\FloatBarrier
\subsubsection{Covariate balance}

\begin{figure}[H]
\includegraphics[width = \textwidth]{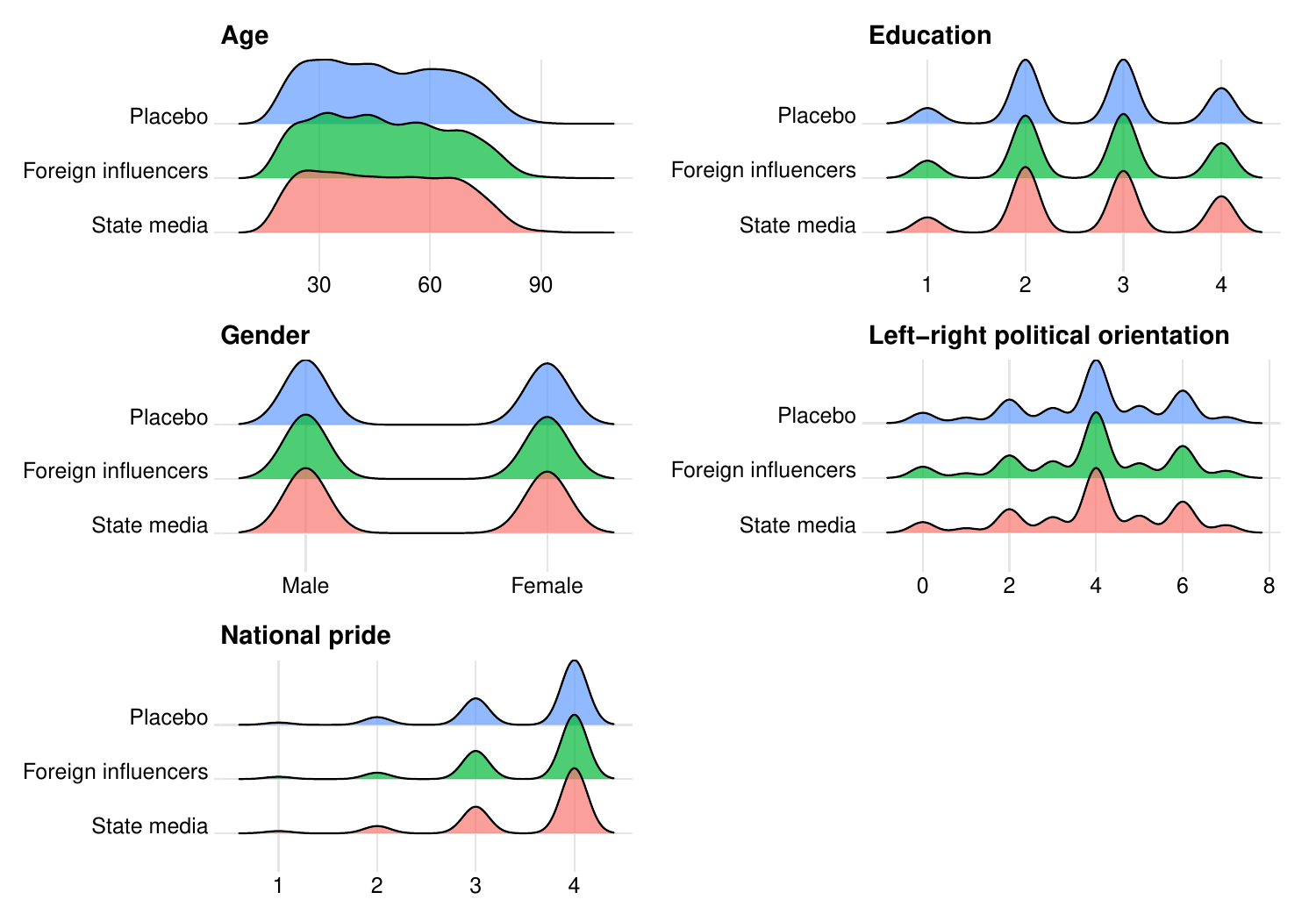}
\caption{Distributions of covariates by treatment group}
\small
\vspace{-0.3cm}
\label{fig: covs_dist}
\end{figure}

\input{tables/balance.tex}

\input{tables/balance_all.tex}

\clearpage
\FloatBarrier
\subsubsection{Local average treatment effects}

\begin{figure}[H]
\includegraphics[width = \textwidth]{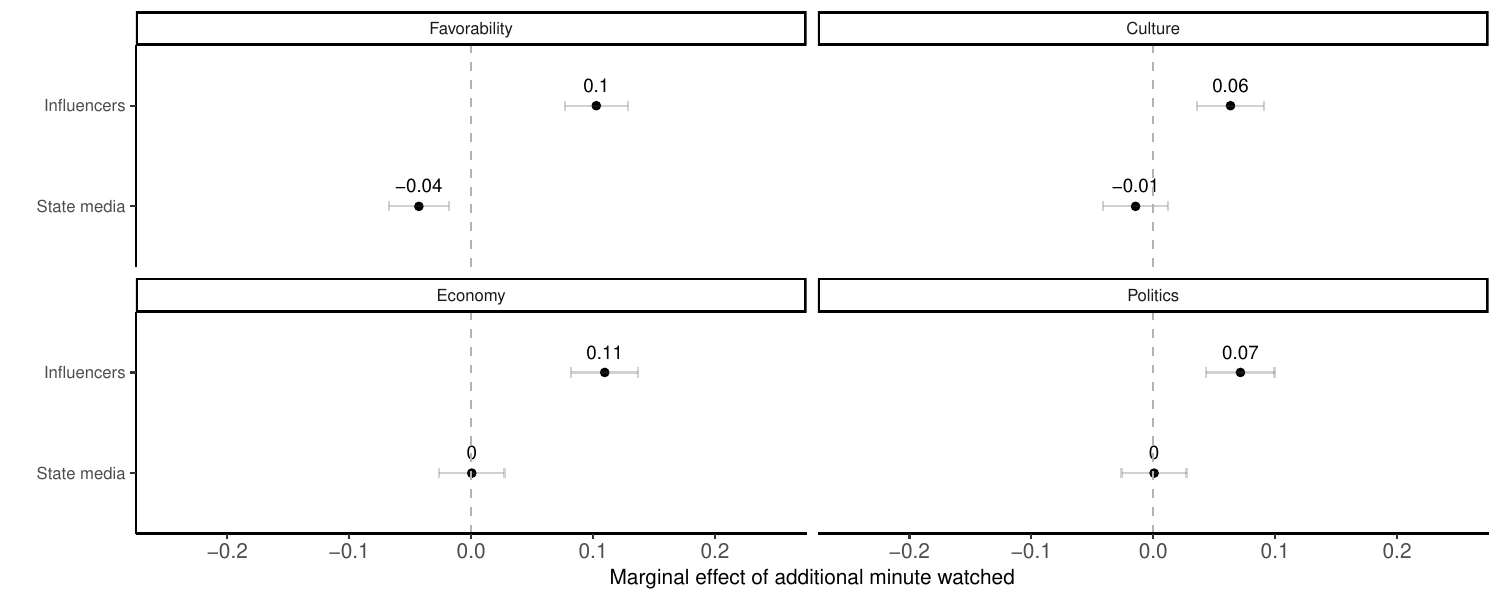}
\caption{Marginal effect of additional minute  of influencer and state media videos watched on attitudes towards China, without covariate adjustment}
\small
\vspace{-0.3cm}
\label{fig: late_nocovs}
\end{figure}

\clearpage
\FloatBarrier
\subsubsection{Secondary outcomes}

\begin{figure}[H]
\includegraphics[width = \textwidth]{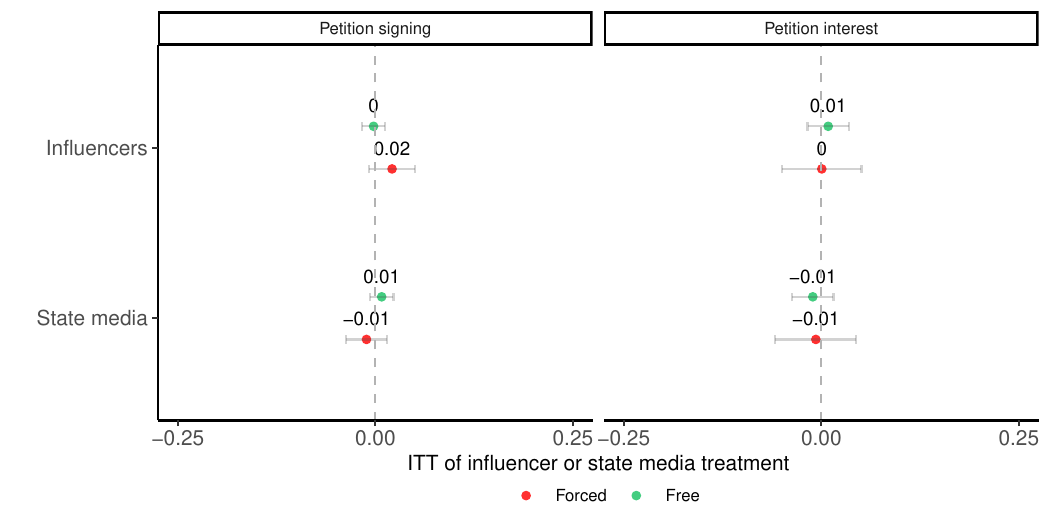}
\caption{Intent-to-treat effects of influencer and state media videos on probability of signing petition}
\small
\vspace{-0.3cm}
\label{fig: petition}
\end{figure}

\begin{figure}[H]
\includegraphics[width = \textwidth]{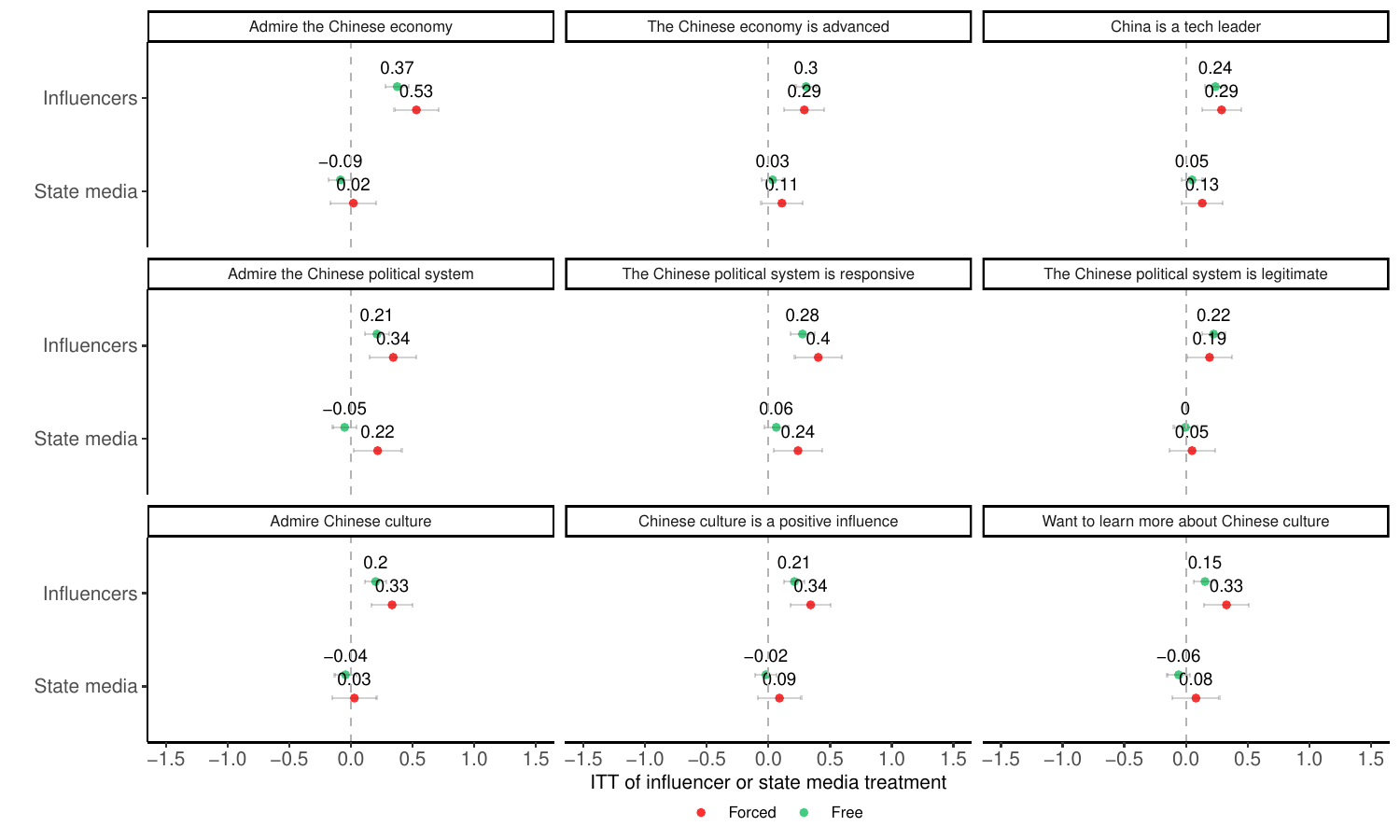}
\caption{Intent-to-treat effects of influencer and state media videos on views about Chinese culture, economy, and politics (non-dimension reduced)}
\small
\vspace{-0.3cm}
\label{fig: all_secondary}
\end{figure}

\begin{figure}[H]
\includegraphics[width = \textwidth]{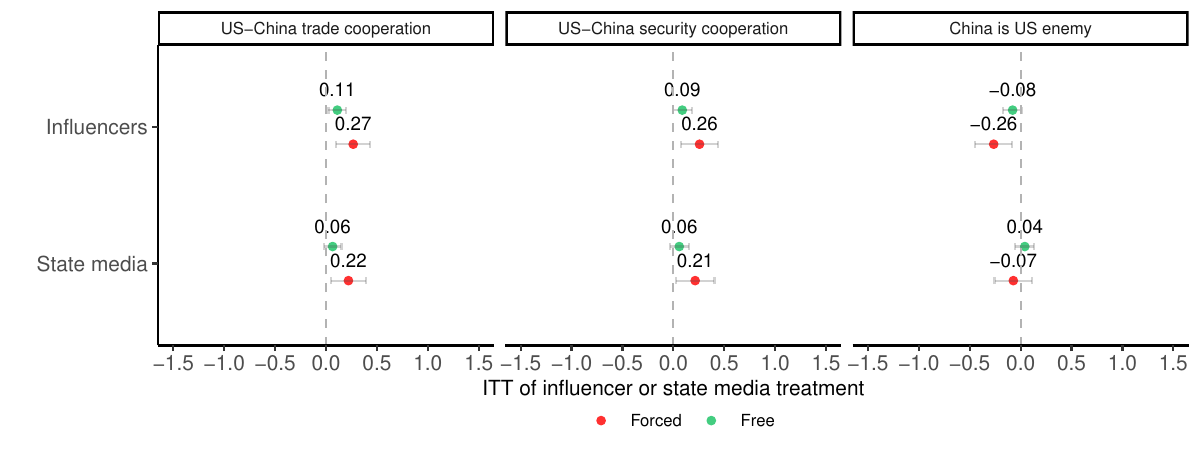}
\caption{Intent-to-treat effects of influencer and state media videos on foreign policy attitudes towards China}
\small
\vspace{-0.3cm}
\label{fig: foreign_policy}
\end{figure}

\begin{figure}[H]
\includegraphics[width = \textwidth]{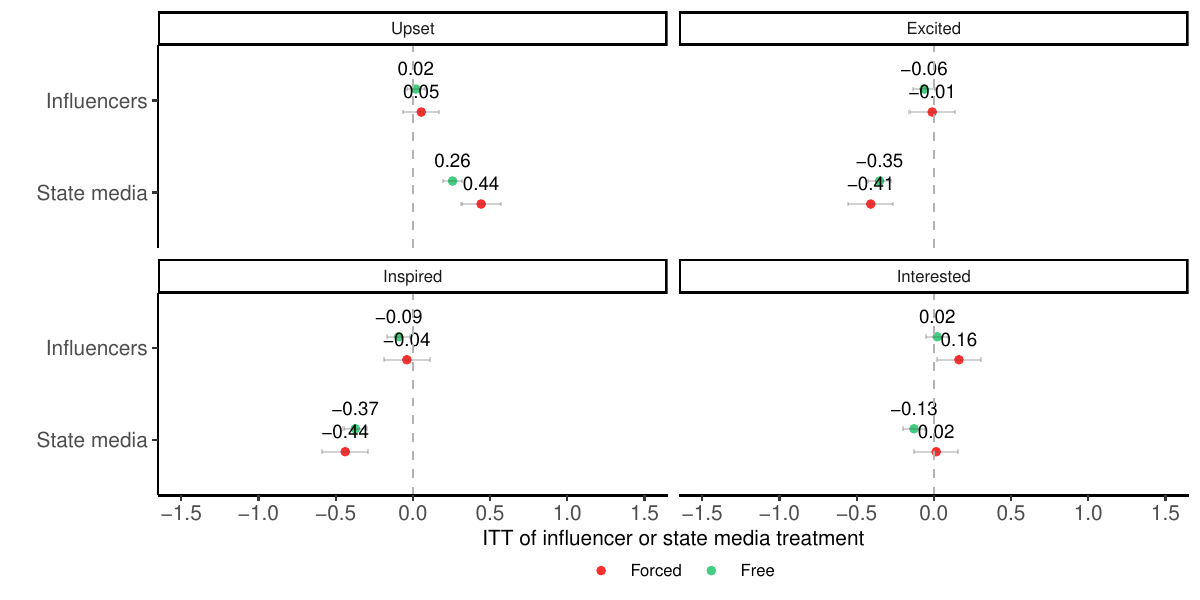}
\caption{Intent-to-treat effects of influencer and state media videos on emotions}
\small
\vspace{-0.3cm}
\label{fig: emotions}
\end{figure}

\clearpage
\FloatBarrier
\subsubsection{Heterogeneous treatment effects}

\begin{figure}[H]
\includegraphics[width = \textwidth]{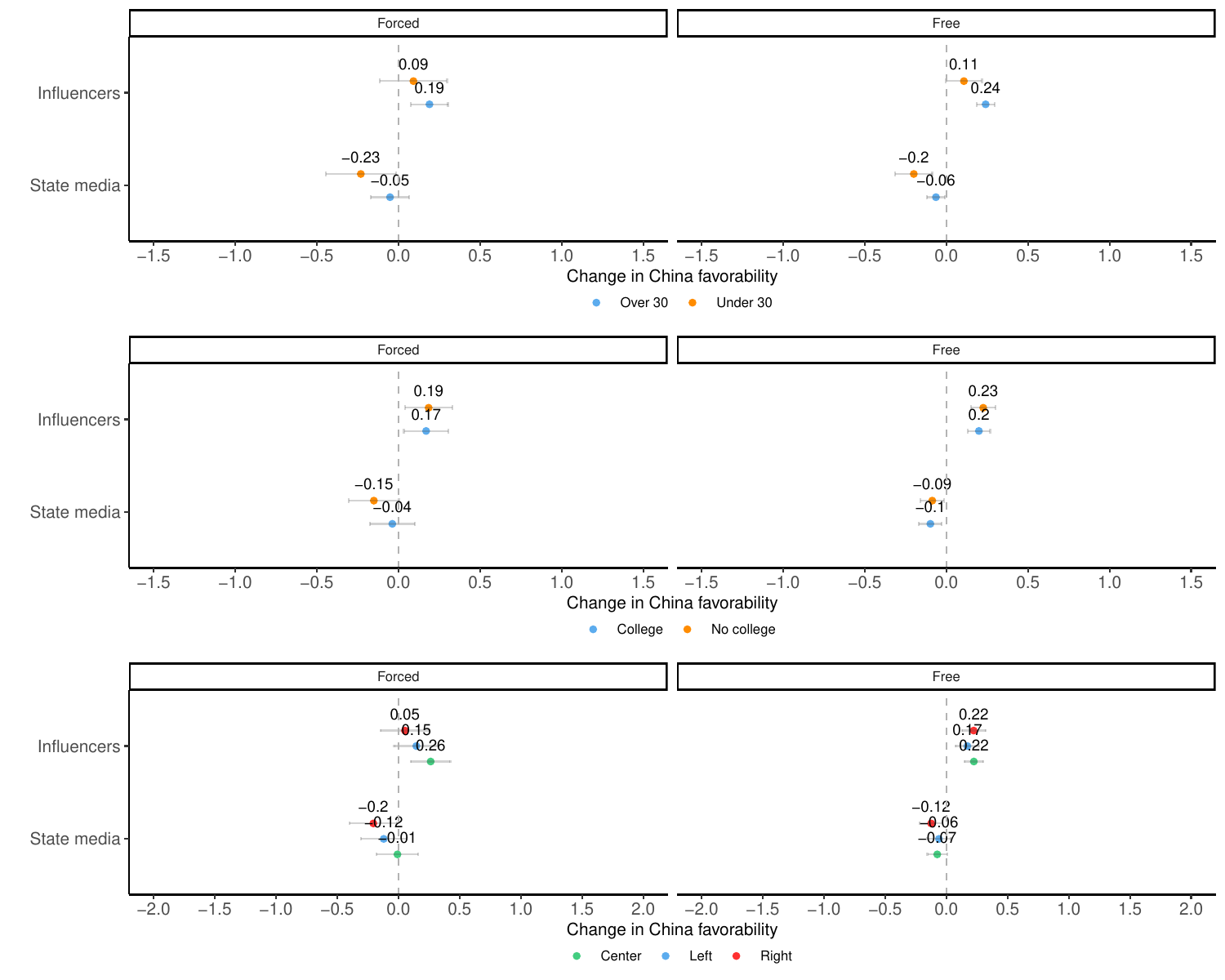}
\caption{Treatment effect heterogeneity: age, education, and left-right political orientation}
\small
\vspace{-0.3cm}
\label{fig: hte}
\end{figure}

\newpage
\subsection{Video-level estimates}

\input{tables/most_positive_videos}

\input{tables/most_negative_videos}

\clearpage
\FloatBarrier
\subsection{Tabular results}

\input{tables/primary_table}

\input{tables/secondary_table_forced}

\input{tables/secondary_table_free}

\clearpage

\subsection{Estimated Monetary Returns} \label{sec: monitary_returns}

We estimate the monetary returns to the foreign influencers in our sample. TikTok requires monetized accounts to be based in the United States, United Kingdom, Germany, Japan, South Korea, France, or Brazil, to have over 10,000 followers, and to have received over 100,000 views in the previous thirty days.\footnote{See \url{https://www.tiktok.com/creator-academy/en/article/creator-rewards-program}.} We therefore filter to accounts meeting these criteria, then estimate monthly revenue for each unique account assuming revenue of between \$0.50 to \$1.0 USD per 1,000 views.\footnote{Journalist accounts document payments of between \$0.50 to \$1.0 USD per 1,000 views on TikTok.} 

We estimate that the top pro-China foreign influencer in our sample in terms of engagement is earning revenue between \$62,000 - \$125,000 USD per month. This represents an outlier, however---only 10\% of our sample earns more than \$10,000 per month by the high estimate of \$1 per 1000 views or \$5,000 per month by the low estimate of \$0.50 per 1000 views. Only 45\% of the sample is making more than \$1,000 per month by the high estimate, with median earnings of roughly \$800 USD by the high estimate and \$400 USD by the low estimate. Nonetheless, this represents a total of 21 accounts generating revenue above real median US personal income, and 9 accounts generating more than \$100,000 in annualized income. Note that these estimates are also independent of any potential payments by state actors or other sponsors, which are unobservable.

\begin{figure}[H]
\includegraphics[width = \textwidth]{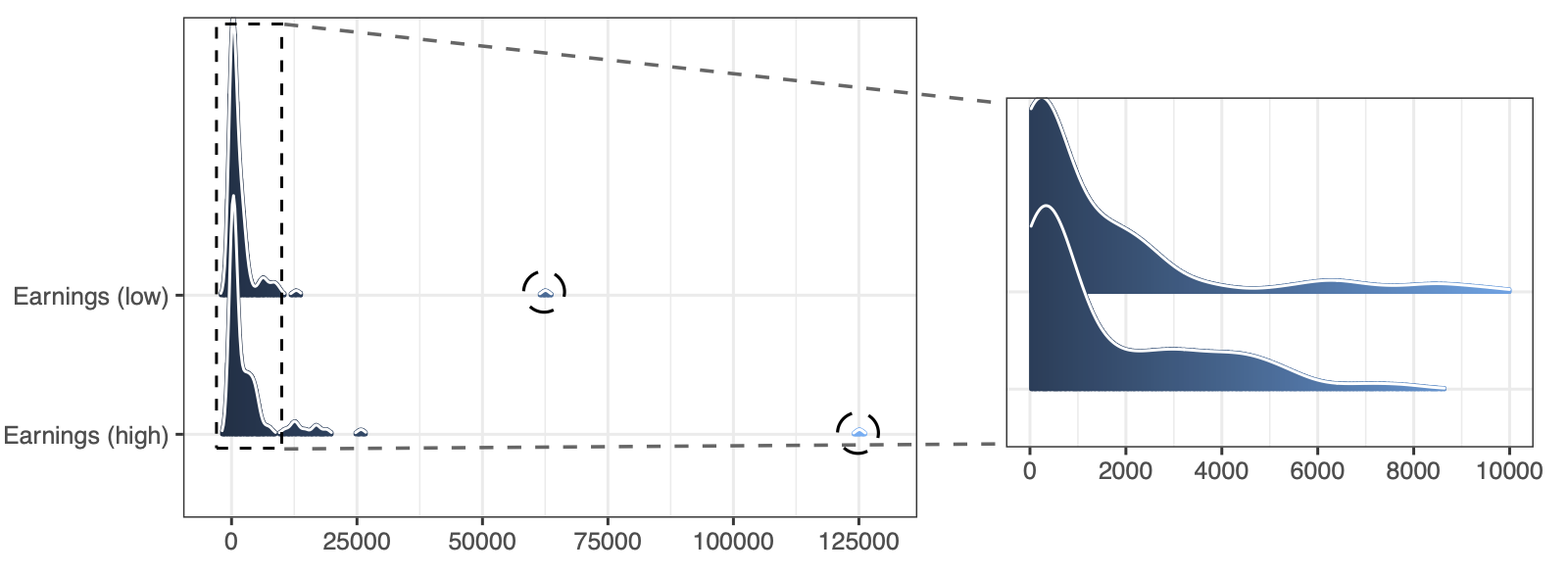}
\caption{Estimated mean monthly earnings of pro-China foreign influencers (USD)}
\small
\vspace{-0.3cm}
\label{fig: earnings}
\end{figure}

\FloatBarrier
\subsection{Robustness}

\input{tables/ri_pvalues}

\input{tables/mcc}

\newpage
\subsection{Compliance}\label{sec: compliance}

The ability to watch or skip pro-China videos in the free choice treatment arms is equivalent to compliance in a one-sided noncompliance framework.  As subjects can choose to watch any duration of pro-China videos they choose, this represents a continuous measure of compliance sometimes referred to as variable treatment intensity. \citet{angrist1995two} show that two-stage least squares (2SLS) applied to a causal model with variable treatment intensity and nonignorable treatment assignment identifies a weighted average of per-unit treatment effects. 

We therefore estimate the weighted average local average treatment effect (LATE)---i.e., the marginal effect of an additional second of watching pro-China videos---from a 2SLS estimation strategy with a continuous measure of treatment intensity defined as the total number of seconds spent watching pro-China videos as our primary compliance-adjusted estimator. Because the marginal effect of each second watched may vary, we will also provide a plot summarizing the effect for each second. This is essentially a binning analysis, which is a common technique to deal with non-linear effects in continuous compliance design, where each bin represents one additional second watched.

We also note that compliance could be defined in a binary or categorical manner. However, our primary definition of compliance will be the continuous measure described above. For a binary definition of compliance in a potential outcomes framework where $d$ refers to treatment status and $z$ refers to treatment assignment, compliance in the free choice treatment groups can be defined as: $d_i(z_i = 1) = 1$ (subject $i$ is assigned to the pro-China treatment and watches \textit{one or more} pro-China videos). Non-compliance can similarly be defined as $d_i(z_i = 1) = 0$ (subject $i$ is assigned to the pro-China treatment but does not watch any pro-China videos). It is also possible to classify compliers in terms of categorical levels of compliance as respondents in the free choice group may choose to watch anywhere between $0$ to $N$ pro-China videos. We can therefore also alter our definition of compliance to $d_i(z_i \in \{1, 2, \ldots, N\}) $ where $d_i(z_i = 1)$ now refers to subject $i$ is assigned to the pro-China treatment and watches \textit{1} pro-China video, $d_i(z_i = 2)$ refers to subject $i$ is assigned to the pro-China treatment and watches \textit{2} pro-China videos, and $d_i(z_i = 3)$ refers to subject $i$ is assigned to the pro-China treatment and watches \textit{3} pro-China videos, and so on.

In all of the approaches above, we calculate complier average causal effects using 2 stage least squares (2SLS) where treatment assignment $Z$ is an instrument for treatment receipt $D$. Additionally, in all of the approaches described above, the exclusion restriction assumption in this setting is that random assignment of pro-China videos is only correlated with our dependent variables through actually \textit{watching} the pro-China videos.

\newpage
\subsection{Debrief}\label{sec: debrief}

Following the post-treatment survey, respondents receive the following debrief: \\

\noindent ``The purpose of this survey was to measure whether watching videos produced by Chinese state-run media and/or TikTok influencers is effective at shifting American attitudes towards the People’s Republic of China, its government, and the Chinese Communist Party. For the purposes of comparison, some respondents watched videos unrelated to China, while others watched videos related to China. \\

\noindent If you watched videos about China, the creators of the video you watched may be independent or they may have been paid or otherwise supported by the Chinese government to produce content that portrays China in a favorable light. Other videos were produced by Chinese state media. Content produced by Chinese state media were labeled as such. \\

\noindent These videos do not represent the opinions of the survey researchers. Moreover, there are alternative views to those presented in the video, views that are more critical of the Chinese government and the Chinese political system and economy. \\

\noindent If you are interested in reading alternative more critical viewpoints on China’s political system, you may find them at the website for Amnesty International (\url{https://www.amnesty.org/en/location/asia-and-the-pacific/east-asia/china/}) and the website for the United States Congressional-Executive Commission on China (\url{https://www.cecc.gov/}). \\

\noindent If you are interested in alternative views on the Chinese economy, you may find them on the website for the Organization for Economic Co-operation and Development (OECD). (\url{https://www.oecd.org/en/topics/sub-issues/economic-surveys/china-economic-snapshot.html}). \\

\noindent part of the survey you were also asked if you would like to sign a petition by Amnesty International. The survey does not have any affiliation with the researchers and does not necessarily represent the opinions of the researchers. The purpose of this question was to measure whether viewing pro-China videos produced by TikTok influencers makes individuals less likely to take political action on topics related to China. While we captured data on whether or not you clicked the link, we do not have access to any data you provided to Amnesty including your name or other identifying information. We also collected data on whether or not you clicked links to external websites to read alternative viewpoints for the purposes of assessing how often individuals chose to view such material.''

%%%%%%%%%%%%%%%%%%%%%%%%%%%%%%%%%%%%%%%%%%%%%%%%%%%%%%%%%%%
% PRE-ANALSYSIS PLAN
%%%%%%%%%%%%%%%%%%%%%%%%%%%%%%%%%%%%%%%%%%%%%%%%%%%%%%%%%%%

\newpage
\section{\centering Pre-analysis plan: \\ 
Foreign influencer operations: How TikTok shapes American perceptions of China} \label{sec: preanalysis}
\vspace{0.5cm}

We measure the effect of pro-China social media influencer content on attitudes and behavior using an experimental design. We first expose participants to a baseline survey, then randomly assign participants to either: \\ 

\noindent (1) A group that is required to watch 4 minutes TikTok videos to which they are exposed. \\

\noindent (2) A group that is exposed to 4 minutes  of TikTok videos, and is free to watch or skip any videos they choose, as they would on the real TikTok platform. \\

\noindent Within group (1), respondents will then be randomly assigned to a random sample of videos from Chinese state media accounts, videos from pro-China influencer accounts, or entertainment-related placebo videos unrelated to China. Within the pro-China videos, videos will be randomly assigned with equal probability on the subjects of politics, economics, and culture, from a larger corpus with equal probability of assignement. Respondents in this group \textit{must watch all of the videos} (i.e., there is no option to skip videos). \\

\noindent Within group (2), respondents will also be randomly assigned to a random sample of videos from Chinese state media accounts, videos from pro-China influencer accounts, or entertainment-related placebo videos unrelated to China. Within the pro-China videos, videos each will be randomly assigned with equal probability from a larger corpus of videos on the subjects of politics, economics, and culture, with equal probability of assignment. However, respondents in this group \textit{can freely choose to watch or skip any of the videos}, but will be required to spend at 4 minutes before proceeding. \\

\noindent Following treatment assignment and exposure, respondent will complete an endline survey. \\

\begin{figure}[H]
\includegraphics[width = \textwidth]{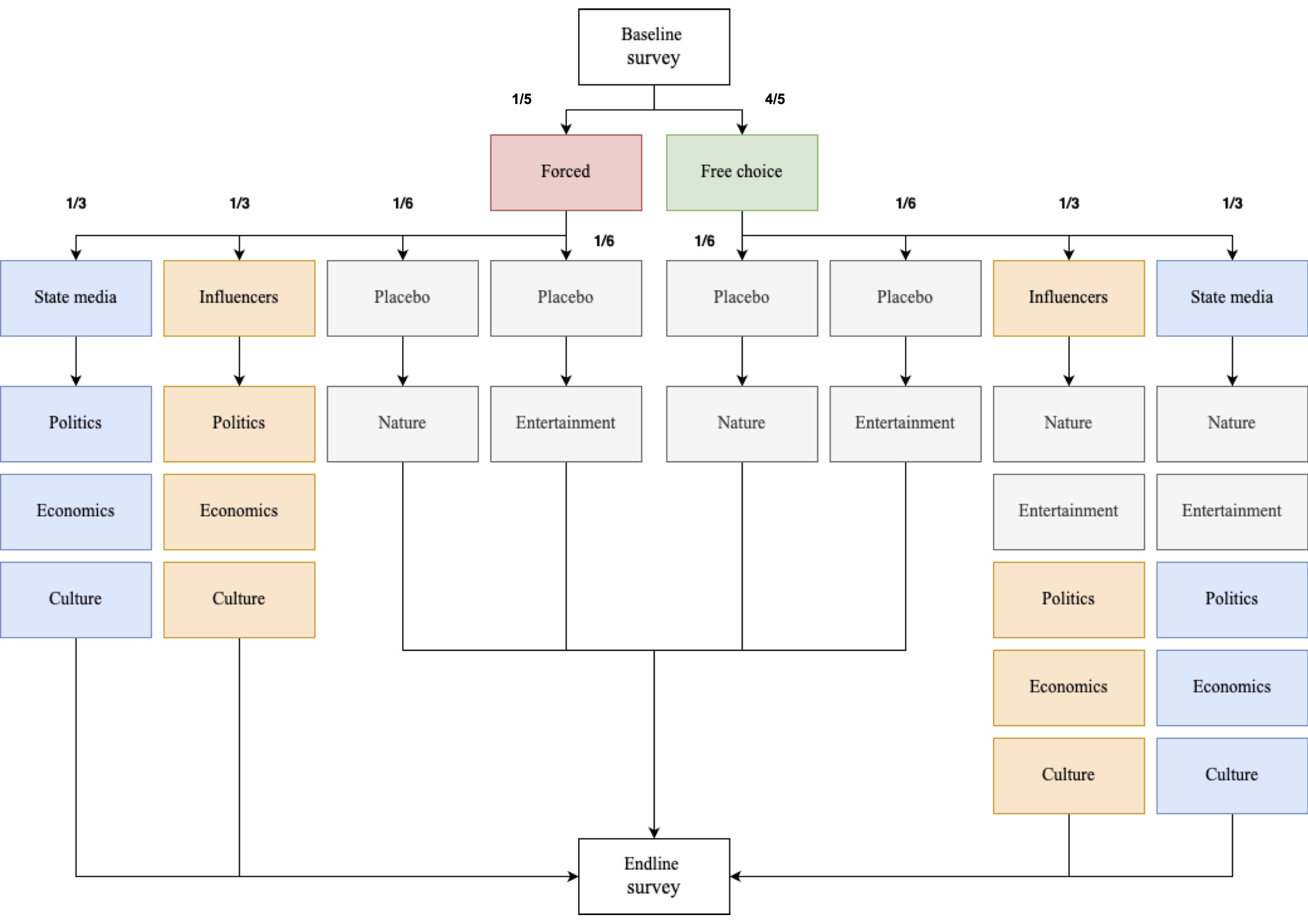}
\caption{Overview of experimental design. }
\small
\vspace{-0.3cm}
\label{figures/fig: experiment_overview_pap}
\singlespace
\noindent
\footnotesize
\textit{Note}: Fractions indicate probability of assignment at each branch. 
\end{figure}

\noindent A visual overview of the experimental design can be found in \autoref{figures/fig: experiment_overview_pap}. \\

\noindent This pre-analysis plan was originally registered prior to collection of data for a pilot study on 500 respondents. In response to findings from the pilot study, we made the following changes: (a) adding the factor analysis outlined below; (b) adding the continuous LATE/2SLS estimation strategy outlined below; (c) adding a series of questions on partisan identification; (d) changing the probabilities of assignment to the forced and free choice groups to increase sample size allocated to the free choice group; (e) updating the power analysis to reflect effect sizes from the pilot study.  We then filed an updated pre-analysis plan before additional data collection. As treatments did not change between the pilot and full study, we will conduct our analysis both including and excluding the results of the pilot study, with results excluding the pilot as robustness checks. 
%\begin{figure}[H]
%\includegraphics[width = \textwidth]{experiment_diagram.png}
%\caption{Overview of experimental design. }
%\small
%\vspace{-0.3cm}
%\label{fig: experiment_overview}
%\singlespace
%\noindent
%\footnotesize
%\end{figure}

\newpage
\subsection{Treatments}

Our treatments take the form of pro-China TikTok videos created by influencers, pro-China TikTok videos created by official state media, or entertainment-related control videos unrelated to China. We will draw from a total corpus of approximately 150 China-related videos that will be randomly assigned to respondents. Within this corpus, we pre-classify whether the videos are cultural, political, or economic in nature, and randomly with equal probability across categories.  %and use these classifications for block random assignment. 

\subsection{Treatment assignment}

Respondents will be recruited from Cint Thoerem, and will be balanced on age, gender, ethnicity, and region. Respondents will then be randomly assigned to one of the two treatment groups (\textit{Forced} or \textit{Free choice})  with 20\% probability of assignment to the \textit{Forced} group and 80\% to the \textit{Free Choice} group).\footnote{We randomly assign more respondents to the \textit{Free Choice} group as we expect smaller treatment effects from this group, and therefore wish to increase sample size and therefore power in this group. See the power analysis below for justification of the selected sample sizes in each condition.} Respondents are then randomly assigned to one of three conditions (\textit{Placebo}, \textit{State Media}, or \textit{Influencers}) within these groups with equal probability. Finally, within the \textit{State Media} and \textit{Influencer} groups, respondents will be randomly assigned political, economic, and culture videos with equal probability of assignment. In short, at each step there will be equal probability of assignment using complete random assignment. See below for an overview of the assignment steps. 

\begin{enumerate}
\item \textit{Forced}: 

\begin{enumerate}
\item[1a.] \textit{Placebo}: Randomly assigned to view either (1) nature or (2) viral entertainment videos with equal probability.
\item[1b.] \textit{State Media}: Randomly assigned to view Chinese state media videos. 
\begin{enumerate}
\item Among Chinese state media videos, randomly assigned politics, economics, and culture videos with equal probability of assignment. 
\end{enumerate}
\item[1c.] \textit{Influencers}: Randomly assigned to view pro-China influencer media videos. 
\begin{enumerate}
\item Among pro-China influencer videos, randomly assigned political, economic, and cultural videos with equal probability of assignment.  
\end{enumerate}
\end{enumerate}

\item \textit{Free choice}: 
\begin{enumerate}
\item[2a.] \textit{Placebo}: Randomly assigned to view either (1) nature or (2) viral entertainment videos with equal probability.
\item[2b.] \textit{State Media}: Randomly assigned to view nature, entertainment and Chinese state media videos. Can skip videos at will.
\begin{enumerate}
\item Among Chinese state media videos, randomly assigned political, economic, and cultural videos with equal probability of assignment. 
\end{enumerate}
\item[2c.] \textit{Influencers}: Randomly assigned to view nature, entertainment and pro-China influencer videos.  Can skip videos at will.
\begin{enumerate}
\item Among pro-China influencer videos, randomly assigned political, economic, and cultural videos with equal probability of assignment. 
\end{enumerate}
\end{enumerate}
\end{enumerate}

\noindent An overview of the experimental design can be found in the figure below. 

\clearpage
\subsection{Compliance}

The ability to watch or skip pro-China videos in the free choice treatment arms is equivalent to compliance in a one-sided noncompliance framework.  As subjects can choose to watch any duration of pro-China videos they choose, this represents a continuous measure of compliance sometimes referred to as variable treatment intensity. \citet{angrist1995two} show that two-stage least squares (2SLS) applied to a causal model with variable treatment intensity and nonignorable treatment assignment identifies a weighted average of per-unit treatment effects. 

We therefore pre-register the weighted average local average treatment effect (LATE)---i.e., the marginal effect of an additional second of watching pro-China videos---from a 2SLS estimation strategy with a continuous measure of treatment intensity defined as the total number of seconds spent watching pro-China videos as our primary compliance-adjusted estimator. Because the marginal effect of each second watched may vary, we will also provide a plot summarizing the effect for each second. This is essentially a binning analysis, which is a common technique to deal with non-linear effects in continuous compliance design, where each bin represents one additional second watched.

We also note that compliance could be defined in a binary or categorical manner. However, our primary definition of compliance will be the continuous measure described above. For a binary definition of compliance in a potential outcomes framework where $d$ refers to treatment status and $z$ refers to treatment assignment, compliance in the free choice treatment groups can be defined as: $d_i(z_i = 1) = 1$ (subject $i$ is assigned to the pro-China treatment and watches \textit{one or more} pro-China videos). Non-compliance can similarly be defined as $d_i(z_i = 1) = 0$ (subject $i$ is assigned to the pro-China treatment but does not watch any pro-China videos). It is also possible to classify compliers in terms of categorical levels of compliance as respondents in the free choice group may choose to watch anywhere between $0$ to $N$ pro-China videos. We can therefore also alter our definition of compliance to $d_i(z_i \in \{1, 2, \ldots, N\}) $ where $d_i(z_i = 1)$ now refers to subject $i$ is assigned to the pro-China treatment and watches \textit{1} pro-China video, $d_i(z_i = 2)$ refers to subject $i$ is assigned to the pro-China treatment and watches \textit{2} pro-China videos, and $d_i(z_i = 3)$ refers to subject $i$ is assigned to the pro-China treatment and watches \textit{3} pro-China videos, and so on.

In all of the approaches above, we calculate complier average causal effects using 2 stage least squares (2SLS) where treatment assignment $Z$ is an instrument for treatment receipt $D$. Additionally, in all of the approaches described above, the exclusion restriction assumption in this setting is that random assignment of pro-China videos is correlated with the treatment of \textit{watching} pro-China videos, and is only correlated with our dependent variables through actually \textit{watching} the pro-China videos.

%We note that a potential threat to unbiased estimation of complier average causal effects in this setting is differential rates of compliance between pro-China and placebo videos. We will therefore test for differential rates of compliance by estimating the average treatment effect of being assigned pro-China videos on view rates. 

%%%%%%%%%%%%%%%%%%%%%%%%

% \subsection{Project overview}

% \noindent
% Drawing on the corpus of videos we collect from the descriptive part of the project, the experiment will take the following steps:

% \begin{enumerate}

%   \item  We identify the most viewed pro-China videos on TikTok associated with different hashtags created by influencer accounts by web-scraping TikTok according to the procedures described above.  
  
%   \item A nationally representative sample of [N] respondents within the United States will be selected for respondent recruitment.
  
%   \item A pre-treatment survey will be administered to all respondents. See [appendix X] for the text of the survey. 
  
%   \item Respondents will be randomly assigned to different treatment groups as described below, and asked to use a simulated version of TikTok. 
  
%   \item Post-random assignment survey instrument responses will be collected (see \nameref{sec: outcomes} for details). 
  
%   \item Results will be analyzed (see \nameref{sec: estimation} for details).
  
%   \item Heterogenous effects analysis will be performed for the variables listed in \nameref{sec: hte}.

% \end{enumerate}

\clearpage
\subsection{Hypotheses}

We will test the following core hypotheses:

\begin{itemize}
\item \textit{Hypothesis 1}: Viewing videos produced by pro-China influencers and state media will increase respondent affinity for China relative to the placebo condition. (See the key outcome question in \nameref{sec: outcomes}).
\item \textit{Hypothesis 2}: Viewing videos produced by pro-China influencers will cause larger shifts in affinity for China than viewing videos produced by Chinese state media. (See the key outcome question in \nameref{sec: outcomes}).
\item \textit{Hypothesis 3}: In the free-choice arm, compliance for those assigned to pro-China influencers will be higher than compliance for those assigned to watch state media. 
\end{itemize}

We also test the following secondary hypotheses:
\begin{itemize}
\item \textit{Hypothesis 3a}: On average, respondents in the forced viewing group will increase affinity for China more than respondents in the free choice group \nameref{sec: outcomes}).
\item \textit{Hypothesis 3b}: Compliers in the free choice group will increase affinity for China less than respondents in the forced choice group (based on the outcome variables defined in \nameref{sec: outcomes}).
\item \textit{Hypothesis 4}:  Respondents assigned to watch pro-China influencers in the forced choice arm will have higher positive affect than those assigned to watch Chinese state media (based on the outcome variables defined in \nameref{sec: outcomes}).
\item \textit{Hypothesis 5}: Videos produced by Chinese state media will shift American respondents' views on China's political system, on the Chinese economy, and U.S.-China policy issues toward those preferred by the Chinese state. Viewing Chinese state media will have no effect on the question asking respondents to sign the petition memorializing the 1989 protests.  (See outcome variables defined in \nameref{sec: outcomes}). Policy-related videos will be more effective at this than cultural videos. 
\item \textit{Hypothesis 6}: Videos produced by Chinese influencers will shift American respondents' views on China's economy but will have null effects on attitudes to the Chinese political system and U.S.-China policy issues towards those preferred by the Chinese state. (See outcome variables defined in \nameref{sec: outcomes}). Viewing pro-China influencers will have no effect on the question asking respondents to sign the petition memorializing the 1989 protests.  Policy-related videos will be more effective at this than cultural videos. 
\item \textit{Hypothesis 7}: Attitudes will shift more for younger respondents and less educated respondents when compared to older respondents and higher educated respondents. 
\item \textit{Hypothesis 8}: Influencer videos will cause respondents to feel inspired, relative to placebo. State media videos will cause respondents to feel upset. 
\item \textit{Hypothesis 9}: Respondents to the center and right will be more likely to be persuaded by influencer videos than respondents to the left.
\end{itemize}

\clearpage
\subsection{Outcomes} \label{sec: outcomes}

We examine the following primary outcome variables of interest:\\

\noindent \textbf{Primary outcome}: China favorability
\begin{itemize}
\item Do you have a very favorable, somewhat favorable, somewhat unfavorable, or very unfavorable opinion of China?
%\item How much confidence do you have in Xi Jinping to do the right thing in world affairs?
\end{itemize}

\noindent \textbf{Behavioral measure}: willingness to take political action related to China
\begin{itemize}
\item In June 1989, following student-led protests, hundreds of anti-government protesters and citizens were killed by People's Liberation Army troops. Would you like to sign a petition memorializing the Tiananmen incident and condemning the Chinese government? This petition is run by Amnesty International and is not affiliated with the researchers. [Yes, at the end of the survey, please take me to the petition. / No thank you, at the end of the survey, please do not take me to the petition.]

We define the behavioral outcome as a binary indicator equal to 1 if a respondent both indicates that they would like to be taken to the petition \textit{and} clicks on the link to the petition, and 0 otherwise. 

\end{itemize}

\noindent \textbf{Secondary outcomes}: attitudes towards China \\ \\
\noindent Attitudes towards China's economy:
\begin{itemize}
\item  To what extent do you agree with the following statement? I admire the Chinese economy.
\item  To what extent do you agree with the following statement? China has an advanced economy.
\item  To what extent do you agree with the following statement? China is a world leader in technology.
\end{itemize}

\noindent Attitudes towards China's political system:
\begin{itemize}
\item  To what extent do you agree with the following statement? I admire the Chinese political system.
\item  To what extent do you agree with the following statement? The Chinese political system is legitimate.
\item  To what extent do you agree with the following statement? The Chinese political system is responsive to the needs of the Chinese people.
\end{itemize}

\noindent Attitudes towards Chinese culture:
\begin{itemize}
\item  To what extent do you agree with the following statement? I admire Chinese culture.
\item  To what extent do you agree with the following statement? Chinese culture has had a positive influence on the world.
\item  To what extent do you agree with the following statement? I am interested in learning more about Chinese culture.
\end{itemize}

\noindent Attitudes towards China-U.S. Policy
\begin{itemize}
%\item To what extent do you agree with the following statement? The United States government should advocate for the overthrow of the Chinese Communist Party.
%\item On balance, do you think of China as a partner of the U.S., a competitor of the U.S. or an enemy of the U.S.?
\item To what extent do you agree with the following statement? China is an enemy of the United States.
\item To what extent do you agree with the following statement? The United States should cooperate closely with China on trade issues.
\item To what extent do you agree with the following statement? The United States should cooperate closely with China on security issues.
\end{itemize}

\noindent \textbf{Mechanism outcomes}: \\ \\
\noindent Questions measuring negative affect and positive affect from the PANAS scale.
\begin{itemize}
\item Please indicate how strongly you are feeling the following emotions: 
\begin{enumerate}
\item  Inspired (Not at all, a little, somewhat, rather strong, extremely strong
\item  Upset (Not at all, a little, somewhat, rather strong, extremely strong
\item  Interested (Not at all, a little, somewhat, rather strong, extremely strong
\item  Excited (Not at all, a little, somewhat, rather strong, extremely strong \\
\end{enumerate}
\noindent Analysis of open-ended responses including topic modeling approaches and qualitative scoring.
\end{itemize}

\noindent Our secondary outcomes will be combined into three indices---economy, politics, and culture---using factor analysis. The code used to conduct the factor analysis is provided below. Factor analysis will not be applied to the foreign policy attitudes questions. Multiple comparisons corrections will be applied to the secondary and exploratory outcomes using the Bonferroni, Holm-Bonferroni, and Benjamin-Hochburg procedures. \\

\noindent \textbf{Factor analysis function}:

\begin{verbatim}
  create_factor <- function(data, dv_names, verbose = TRUE) {
  
  # Keep variables of interest and ensure that they're numeric 
  sub <- data[, dv_names] %>% mutate_all(funs(as.numeric(as.character(.))))
  
  # Impute missing values with median
  impute <- sub 
  impute <- sapply(impute, function(x) ifelse(is.na(x), median(x, na.rm = T), x))
  
  # Principal component, scaled to have mean 0/sd 1
  f <- princomp(impute, cor = TRUE) 
  if (verbose) print(loadings(f))
  dv <- f$scores[, 1]
  dv <- as.numeric(scale(dv))
  
  # Make sure the variable points the correct way ()
  if (cor(dv, data$outcome, use = "complete") < 0) dv <- -1 * dv  
  
  # If a row in the original data has more than 50% NAs, then replace the score
  # with NA
  bool <- apply(sub, 1, function(x) sum(is.na(x)) / ncol(sub) > 0.5)
  dv[bool] <- NA
  dv
  
}
\end{verbatim}

\clearpage
\subsection{Estimation procedures} \label{sec: estimation}

Our primary estimands are: (1) the average treatment effect of watching a pro-China influencer video on the outcomes listed above when compared to the Placebo group, (2) the average treatment effect of state media videos on the outcomes listed above when compared to the Placebo group, and (3) the complier average causal effect of watching a pro-China influencer video on the outcomes listed above when compared to the Placebo group, and (4) the complier average causal effect of watching a pro-China state media video on the outcomes listed above when compared to the Placebo group. 

We will also compare effect sizes among those in the free choice group and forced choice group. For the free choice group we will calculate both intent-to-treat (ITT) effects and complier average causal effects.

\subsubsection{Average Treatment Effect}

For those in the forced choice group, we will estimate the following average treatment effects (ATE), and the estimator will include covariate adjustment:

\begin{enumerate}
  \item \textit{State Media} vs. \textit{Placebo} 
  \item \textit{Influencer} vs. \textit{Placebo} 
   \item \textit{Influencer} vs. \textit{State Media} 
\end{enumerate}

We include the following pre-treatment covariates in the regression specification: \textit{gender, age, education, national pride, and left-right political orientation}. In the event of missingness, missing covariates will be imputed using the predictive mean matching method in the MICE package in \textsf{R}. We will compute HC2 robust standard errors.

The ATE will be estimated using the ``lm\_robust'' function in the ``estimatr'' package in R \citep{blair2019declaring}.  The code that will be used is as follows: 

\begin{verbatim}
lm_robust(outcome ~ treatment + covs, data = df)
\end{verbatim}

\noindent where covs is the list of covariates above and country is an indicator for each of the 19 countries. In the event that standard covariate adjustment worsens precision, we will estimate treatment effects using the estimator outlined by \citet{lin2013agnostic} using the code below:

\begin{verbatim}
lm_lin(outcome ~ treatment, covs, data = df)
\end{verbatim}

Results without covariate adjustment will be reported in the appendix. We expect these results to be similar but less precisely estimated due to the exclusion of prognostic covariates. When interpreting the results, we will rely primarily on the covariate-adjusted estimates. We will also calculate randomization inference p-values for the main outcome variables in the appendix.

\subsubsection{Intent-to-treat effects}

For those in the free choice group, we will calculate the same ITTs as the ATEs described above, substituting receipt of treatment for assignment to treatment in the regression specification. We will also compare the ITT in the free choice group to the ATE in the forced choice group. The ITT will be estimated using the ``lm\_robust'' function in the ``estimatr'' package in R:

\begin{verbatim}
lm_robust(outcome ~ treatment_assignment + covs, data = df)
\end{verbatim}

\subsubsection{Local average treatment effects}

We will also estimate local average treatment effects / complier average causal effects (CACE) as described in the ``compliance'' section above. We will calculate the same LATEs as the ITTs/ATEs described above.  We will estimate the LATE using two-stage least squares using the ``iv\_robust'' function in the ``estimatr'' package in R: %In the event that there is not differential compliance between the placebo and treatment groups (i.e., watch rates are the same for both placebo and treatment videos), we will estimate the CACE as the average treatment effect for individuals who watched the videos only (i.e., compliers). 

\begin{verbatim}
iv_robust( 
\end{verbatim} 
\vspace{-0.5cm}
\begin{verbatim} 
outcome ~ treatment_receipt + covs | treatment_assignment + covs, data = df
\end{verbatim}
\vspace{-0.5cm}
\begin{verbatim} 
)
\end{verbatim}

\subsection{Treatment effect heterogeneity} \label{sec: hte}

We will examine the following heterogeneous treatment effects: 

\begin{enumerate}
  \item Age - under 30 vs. over 30. 
  \item Education - college educated vs. non-college educated 
  \item Left-right political alignment - left vs. center vs. right
  \item Partisan identification - Republican vs. Democrat
\end{enumerate}

%\subsection{Heterogeneous treatment effects}

We will examine treatment effect heterogeneity by calculating conditional average treatment effects (CATEs). A CATE is an average treatment effect specific to a subgroup of subjects, where the subgroup is defined by subjects’ attributes. Heterogeneous treatment effects will be estimated by regressing the outcome variables on treatments for each of the subgroups specified in \nameref{sec: hte}. This will be conducted using the ``lm\_robust'' function in the ``estimatr'' package in R.

\begin{verbatim}
lm_robust(outcome ~ treatment, data = df, subset = covariate == "subgroup")
\end{verbatim}

\subsection{Threats to inference}

\subsubsection{Attrition} \label{sec: attrition}

We will examine whether there is differential attrition between the placebo and treatment groups, as well as between the forced choice and free choice groups. We will conclude that there is significant attrition if the difference in the rate of attrition between the groups is statistically significant (p$<$0.05). If there is significant differential attrition, we will present reweighted results and trimming bound results in the paper's appendix, as appropriate, following the procedures outlined in \cite{gerber2012field}. 

\subsubsection{Attention checks}

In  light  of  recent  evidence  of  decreased  attention  in  online  samples \citep{peyton2020generalizability},  respondents will be screened according to pre-treatment attention checks and dropped from the sample of analysis if they fail the attention check. Our attention checks will take the following form: 

\begin{enumerate}
\item ``For our research, careful attention to survey questions is critical! To show that you are paying attention, please select 'I have a question'.''
\item ``People have different tastes in movies. For this question, however, we are not interested in your taste but want to test whether survey takers are reading questions carefully. Below, please select the options ``Romance'' and ``Science Fiction.''''
\end{enumerate}

%However, we will also provide the main results listed in \nameref{sec: estimation} using the full sample (i.e., without dropping those who fail the attention check) in the appendix. We will also examine if attention check failure is correlated with our pre-treatment variables. 

\subsection{Potential Early Survey Stop}

One concern with online samples and our survey partner is the possibility that as the fielding of the survey progresses, the quality of responses may decrease. While fielding the survey, we will monitor the rate at which surveys are completed and the length of survey completion. If we observe a significant slow-down in the rate at which survey takers are completing the survey, we will halt the survey. If we observe a significant decrease in the amount of time it is taking survey takers to complete the survey below a 10 minute baseline, we will halt the survey. At this point we will not analyze the outcome data, but if based on indications like the number of speeders we assess that the survey is no longer receiving high-quality responses, we will end the survey.

\subsection
{Data collection issues}

We note here a minor technical issue that arose during response collection that a affected a number of respondents in the pilot study. Some respondents were assigned to two treatment groups (i.e., they watched two sets of videos) due to internet connection issues and/or refreshing the experiment midway through the study. This affected a total of 9 respondents, who were subsequently removed from the analysis sample. We will also monitor and remove any such respondents for the full study.  

\clearpage
\subsection{Power analysis}

The power analysis below depicts the intent to treat effect of the three treatment groups (Placebo, State Media, and Influencers), separately for the forced and free choice groups. The analysis indicates that due to expectations of larger treatment effects sizes in the forced group, a larger sample should be expended on the free choice group. The power analyses below assume the distributions of potential outcomes / treatment effect sizes among respondent as found in the pilot study and were run using 10,000 simulations. 

\subsubsection{Forced choice}

The power analysis for the forced choice group indicates sufficient power (defined here as the conventional 80\% power) with a sample size of roughly 500 respondents. This is not surprising as significant results were uncovered from the pilot alone in the forced choice group. The distributions of p values across 500 simulations can be found in the figure below. 

% \begin{table}[H]
% \centering
% \resizebox{10cm}{!}{
% \begin{tabular}{lccc}
% \toprule
% Estimate & Power & SE(power) & N \\
% \midrule
% \cellcolor{gray!6}{State media \& influencers vs. placebo} & \cellcolor{gray!6}{0.93} & \cellcolor{gray!6}{0.01} & \cellcolor{gray!6}{500} \\
% \cellcolor{gray!6}{State media vs. placebo} & \cellcolor{gray!6}{0.93} & \cellcolor{gray!6}{0.01} & \cellcolor{gray!6}{500} \\
% \cellcolor{gray!6}{Influencers vs. placebo} & \cellcolor{gray!6}{0.98} & \cellcolor{gray!6}{0.01} & \cellcolor{gray!6}{500} \\
% \cellcolor{gray!6}{State media vs. influencers} & \cellcolor{gray!6}{0.93} & \cellcolor{gray!6}{0.01} & \cellcolor{gray!6}{500} \\
% \bottomrule
% \end{tabular}}
% \end{table}

\begin{figure}[H]
\includegraphics[width = \textwidth]{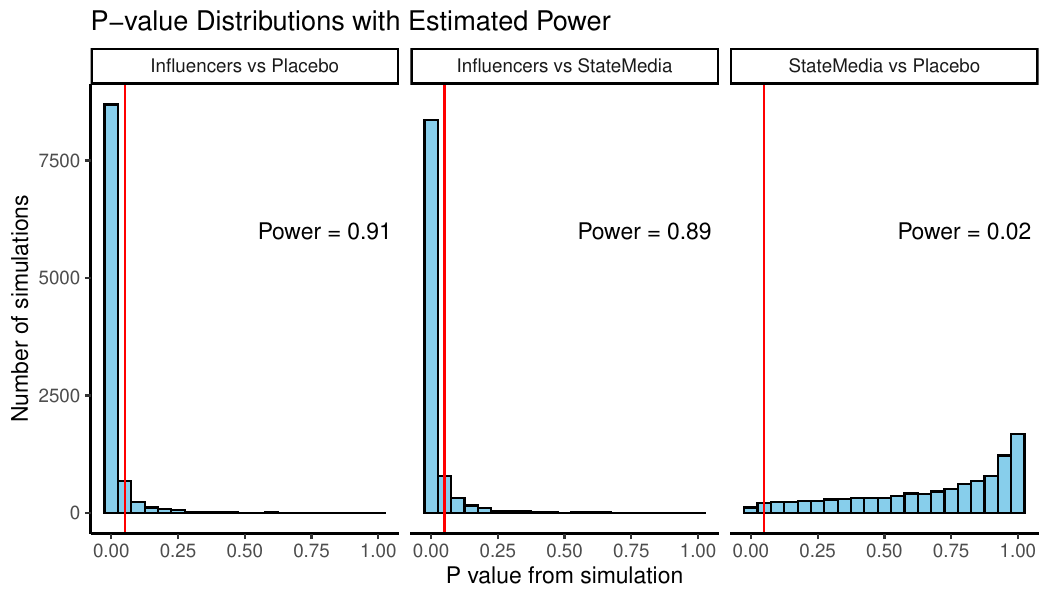}
\caption{Distribution of p values in forced arm (N = 500 respondents)}
\end{figure}

\subsubsection{Free choice}

However, for the free choice group, due to high expected levels of noncompliance and a correspondingly lower ITT, the required sample size to reach 80\% power is significantly larger. Using the distribution of potential outcomes from the pilot, we do not reach 80\% power in each of the treatment arms until there is a sample size of 6000 respondents in the free choice group. 

The distributions of p values across 500 simulations can be found in the figures below. 

\begin{figure}[H]
\includegraphics[width = \textwidth]{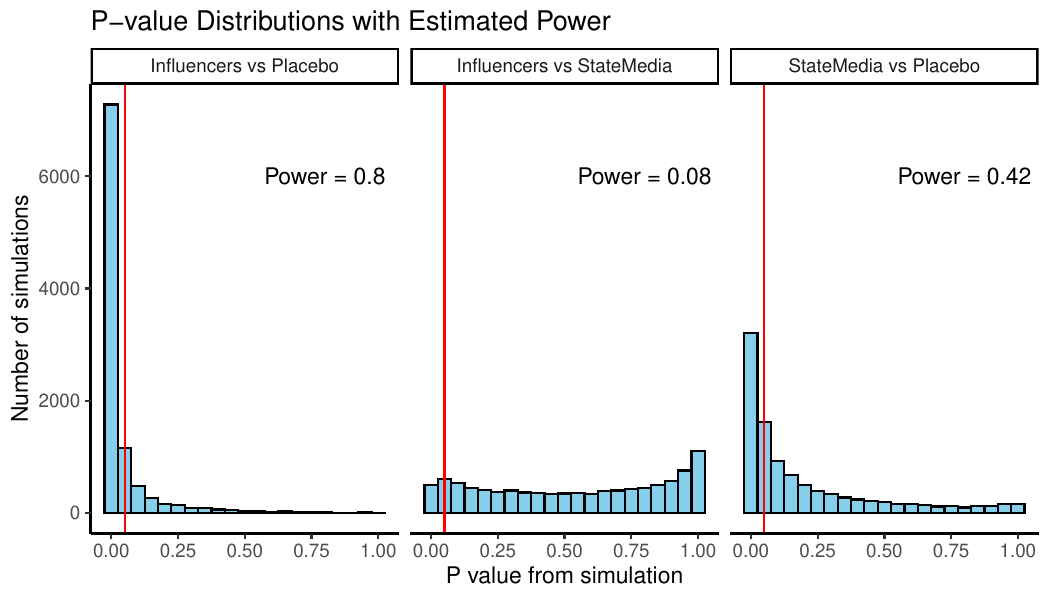}
\caption{Distribution of p values in free choice arm (N = 6000 respondents)}
\end{figure}

These simulations do not cover additional sample size that would likely be necessary to detect differences between the different types of videos or heterogeneous effects by age, education, or pre-existing affinity for China. To ensure sufficient power to detect these effects, additional respondents would be necessary. 
%TC:endignore

\end{document}

%% file: tables/descriptive_table.tex
\begin{table}[ht]
  \centering
  \begin{subtable}[t]{0.45\textwidth}
    \centering
    \begin{tabular}{@{}lll@{}}
\toprule
Account type        & Videos & Accounts \\ \midrule
Contains ``\#China''           & 40,684         & 23,442           \\
Foreign influencers & 3,321          & 204             \\
State media         & 316           & 24              \\ \bottomrule
\end{tabular}
    \caption{Total number of videos and accounts by account type}
  \end{subtable}
  \quad
  \begin{subtable}[t]{0.45\textwidth}
    \centering
    \begin{tabular}{@{}lll@{}}
\toprule
Category       & N accounts & \% of accounts\footnote{Percentages sum to over 100\% as creators may discuss multiple topics.} \\ \midrule
Culture        & 116        & 56                   \\
Economy        & 87         & 42                   \\
Politics       & 48         & 23                           \\ \bottomrule
\end{tabular}
    \caption{Topic prevalence for foreign influencer accounts (N = 204 accounts)}
  \end{subtable}
  \caption{Overview of foreign influencer and state media accounts}
\end{table}

%% file: tables/balance.tex
\begin{table}[H]
\centering
\resizebox{\ifdim\width>\linewidth 0.9\linewidth\else\width\fi}{!}{
\begin{talltblr}[         %% tabularray outer open
caption={Balance table: by treatment group},
note{}={Note: Covariates in balance table are the same as covariates used in covariate adjustment.},
]                     %% tabularray outer close
{                     %% tabularray inner open
colspec={Q[]Q[]Q[]Q[]Q[]Q[]Q[]Q[]Q[]},
cell{1}{1}={}{halign=l, halign=c,},
cell{1}{2}={c=2,}{halign=r, halign=c, halign=c,},
cell{1}{3}={}{halign=r, halign=c,},
cell{1}{4}={c=2,}{halign=r, halign=c, halign=c,},
cell{1}{5}={}{halign=r, halign=c,},
cell{1}{6}={c=2,}{halign=r, halign=c, halign=c,},
cell{1}{7}={}{halign=r, halign=c,},
cell{1}{8}={c=2,}{halign=r, halign=c, halign=c,},
cell{1}{9}={}{halign=r, halign=c,},
cell{2-7}{1}={}{halign=l,},
cell{2-7}{2}={}{halign=r,},
cell{2-7}{3}={}{halign=r,},
cell{2-7}{4}={}{halign=r,},
cell{2-7}{5}={}{halign=r,},
cell{2-7}{6}={}{halign=r,},
cell{2-7}{7}={}{halign=r,},
cell{2-7}{8}={}{halign=r,},
cell{2-7}{9}={}{halign=r,},
}                     %% tabularray inner close
\toprule
& Foreign influencers (N=2868) &  & Placebo (Influencer) (N=1427) &  & Placebo (Nature) (N=1377) &  & State media (N=2767) &  \\ \cmidrule[lr]{2-3}\cmidrule[lr]{4-5}\cmidrule[lr]{6-7}\cmidrule[lr]{8-9}
& Mean & Std. Dev. & Mean & Std. Dev. & Mean & Std. Dev. & Mean & Std. Dev. \\ \midrule %% TinyTableHeader
Gender & \num{0.5} & \num{0.5} & \num{0.5} & \num{0.5} & \num{0.5} & \num{0.5} & \num{0.5} & \num{0.5} \\
Age & \num{47.3} & \num{17.5} & \num{47.1} & \num{17.5} & \num{48.1} & \num{17.5} & \num{47.4} & \num{17.6} \\
Education & \num{2.7} & \num{0.9} & \num{2.7} & \num{0.9} & \num{2.7} & \num{0.9} & \num{2.7} & \num{0.9} \\
National Pride & \num{3.5} & \num{0.7} & \num{3.5} & \num{0.7} & \num{3.5} & \num{0.7} & \num{3.5} & \num{0.7} \\
Political orientation & \num{3.9} & \num{1.7} & \num{3.9} & \num{1.7} & \num{3.9} & \num{1.7} & \num{3.9} & \num{1.7} \\
\bottomrule
\end{talltblr}
}
\end{table}

%% file: tables/balance_all.tex
\begin{table}[H]
\centering
\resizebox{\ifdim\width>\linewidth 0.9\linewidth\else\width\fi}{!}{
\begin{talltblr}[         %% tabularray outer open
caption={Balance table: by forced vs. free choice groups and treatment group},
note{}={Note: Covariates in balance table are the same as covariates used in covariate adjustment.},
]                     %% tabularray outer close
{                     %% tabularray inner open
colspec={Q[]Q[]Q[]Q[]Q[]Q[]Q[]Q[]Q[]Q[]Q[]Q[]Q[]},
cell{1}{1}={}{halign=l, halign=c,},
cell{1}{10}={c=2,}{halign=r, halign=c, halign=c,},
cell{1}{11}={}{halign=r, halign=c,},
cell{1}{12}={c=2,}{halign=r, halign=c, halign=c,},
cell{1}{13}={}{halign=r, halign=c,},
cell{1}{2}={c=2,}{halign=r, halign=c, halign=c,},
cell{1}{3}={}{halign=r, halign=c,},
cell{1}{4}={c=2,}{halign=r, halign=c, halign=c,},
cell{1}{5}={}{halign=r, halign=c,},
cell{1}{6}={c=2,}{halign=r, halign=c, halign=c,},
cell{1}{7}={}{halign=r, halign=c,},
cell{1}{8}={c=2,}{halign=r, halign=c, halign=c,},
cell{1}{9}={}{halign=r, halign=c,},
cell{2-7}{1}={}{halign=l,},
cell{2-7}{10}={}{halign=r,},
cell{2-7}{11}={}{halign=r,},
cell{2-7}{12}={}{halign=r,},
cell{2-7}{13}={}{halign=r,},
cell{2-7}{2}={}{halign=r,},
cell{2-7}{3}={}{halign=r,},
cell{2-7}{4}={}{halign=r,},
cell{2-7}{5}={}{halign=r,},
cell{2-7}{6}={}{halign=r,},
cell{2-7}{7}={}{halign=r,},
cell{2-7}{8}={}{halign=r,},
cell{2-7}{9}={}{halign=r,},
}                     %% tabularray inner close
\toprule
& Placebo (Forced) (N=571) &  & State media (Forced) (N=567) &  & Influencers (Forced) (N=603) &  & Placebo (Free) (N=2233) &  & State media (Free) (N=2200) &  & Influencers (Free) (N=2265) &  \\ \cmidrule[lr]{2-3}\cmidrule[lr]{4-5}\cmidrule[lr]{6-7}\cmidrule[lr]{8-9}\cmidrule[lr]{10-11}\cmidrule[lr]{12-13}
& Mean & Std. Dev. & Mean & Std. Dev. & Mean & Std. Dev. & Mean & Std. Dev. & Mean & Std. Dev. & Mean & Std. Dev. \\ \midrule %% TinyTableHeader
Gender & \num{0.5} & \num{0.5} & \num{0.5} & \num{0.5} & \num{0.5} & \num{0.5} & \num{0.5} & \num{0.5} & \num{0.5} & \num{0.5} & \num{0.5} & \num{0.5} \\
Age & \num{47.3} & \num{17.5} & \num{46.7} & \num{17.3} & \num{46.6} & \num{18.1} & \num{47.7} & \num{17.5} & \num{47.6} & \num{17.7} & \num{47.5} & \num{17.4} \\
Education & \num{2.7} & \num{0.9} & \num{2.7} & \num{0.9} & \num{2.7} & \num{0.9} & \num{2.7} & \num{0.9} & \num{2.6} & \num{0.9} & \num{2.6} & \num{0.9} \\
National Pride & \num{3.5} & \num{0.7} & \num{3.5} & \num{0.7} & \num{3.5} & \num{0.7} & \num{3.5} & \num{0.7} & \num{3.5} & \num{0.7} & \num{3.5} & \num{0.7} \\
Political orientation & \num{3.8} & \num{1.8} & \num{3.8} & \num{1.8} & \num{3.7} & \num{1.8} & \num{3.9} & \num{1.7} & \num{3.9} & \num{1.7} & \num{3.9} & \num{1.7} \\
\bottomrule
\end{talltblr}
}
\end{table}

%% file: tables/most_positive_videos.tex
\begin{table}[htbp]
\centering
\caption{15 Most Positive Influence Videos (Free Choice)}
\begin{adjustbox}{width=\textwidth}
\begin{tabular}{lcccccc}
\toprule
\label{tab:positive_videos}
\textbf{Video Description}& \textbf{Estimate} & \textbf{SD} & \textbf{Conf Int Low} & \textbf{Conf Int High} & \textbf{P Value} & \textbf{Video Type} \\
\midrule
A Chinese influencer describes reasons why there is no homelessness in China & 0.3782 & 0.0726 & 0.2360 & 0.5204 & 1.86E-07 & Influencer  \\
Travel guide to Chengdu that showcases attractions in the area of the city  & 0.3151 & 0.0763 & 0.1656 & 0.4646 & 3.61E-05 & Influencer \\
A video of a family-friendly Chinese electric vehicle   & 0.3050 & 0.0739 & 0.1602 & 0.4498 & 3.65E-05 & Influencer \\
A video that showcases a major avenue in China with Chinese flags & 0.2920 & 0.0749 & 0.1452 & 0.4388 & 9.71E-05 & Influencer \\
An American influencer tours a luxury hotel in a desert in China & 0.2899 & 0.0836 & 0.1261 & 0.4537 & 0.00052 & Influencer \\
An American businessmen discuss how much China outperforms the US in manufacturing & 0.2852 & 0.0908 & 0.1073 & 0.4630 & 0.00168 & Influencer \\
Three women in traditional Chinese clothing perform a traditional Chinese dance & 0.2800 & 0.0709 & 0.1411 & 0.4190 & 7.81E-05 & Influencer \\
A video that showcases Chinese high-speed trains. & 0.2793 & 0.0691 & 0.1438 & 0.4148 & 5.35E-05 & Influencer \\
A Chinese influencer walking in the street in Europe and discussing why the quality of life in China is better than the U.S. & 0.2754 & 0.0721 & 0.1341 & 0.4167 & 0.00013 & Influencer \\
A Chinese influencer in Starbucks in China imitates Donald Trump and eats a Persimmon cake & 0.2729 & 0.0724 & 0.1311 & 0.4148 & 0.00016 & Influencer \\
A video of the city of Shenzhen showing its modern skyline and calm traffic. & 0.2727 & 0.0701 & 0.1353 & 0.4101 & 0.00010 & Influencer \\
A video that showcases a Chinese electric sports car. & 0.2558 & 0.0781 & 0.1027 & 0.4090 & 0.00106 & Influencer \\
A British influencer shows the audience an ultra-modern ``cyberpunk'' Chinese city  & 0.2526 & 0.0846 & 0.0868 & 0.4183 & 0.00282 & Influencer \\
A British influencer discusses China's rapid economic growth intercut with modern cityscapes  & 0.2505 & 0.0921 & 0.0699 & 0.4311 & 0.00655 & Influencer \\
A video of Shanghai skyscrapers at night & 0.2471 & 0.0816 & 0.0871 & 0.4071 & 0.00248 & Influencer \\
\bottomrule
\end{tabular}
\end{adjustbox}
\end{table}

%% file: tables/most_negative_videos.tex
\begin{table}[htbp]
\centering
\caption{15 Most Negative Influence Videos (Free Choice)}
\begin{adjustbox}{max width=\textwidth}
\begin{tabular}{lcccccc}
\toprule
\label{tab:negative_videos}
\textbf{Video Description}& \textbf{Estimate} & \textbf{SD} & \textbf{Conf Int Low} & \textbf{Conf Int High} & \textbf{P Value} & \textbf{Video Type} \\
\midrule
A Chinese policeman attempts to control monkeys on a rural road. & -0.2891 & 0.0798 & -0.4455 & -0.1327 & 0.00029 & Statemedia \\
A short speech by the new Chinese ambassador to the US, declaring his role is to safeguard China's interests. & -0.2643 & 0.0846 & -0.4301 & -0.0986 & 0.00177 & Statemedia \\
Chinese MOFA response to Trump's tariffs. & -0.2460 & 0.0811 & -0.4049 & -0.0871 & 0.00241 & Statemedia \\
The Chinese military marches in Moscow on Victory Day. & -0.2455 & 0.0821 & -0.4064 & -0.0847 & 0.00278 & Statemedia \\
Xi meets Putin in Russia's Victory Day at Moscow while the Chinese military marches.  & -0.2385 & 0.0819 & -0.3991 & -0.0779 & 0.00360 & Statemedia \\
A clip of Trump giving a speech and refusing to mention the names of other senators in the event & -0.2369 & 0.0820 & -0.3977 & -0.0761 & 0.00388 & Statemedia \\
China's MOFA welcomes foreign leaders from Europe to China for a visit.  & -0.2308 & 0.0768 & -0.3812 & -0.0803 & 0.00264 & Statemedia \\
Coverage of the protests against Trump in Washington & -0.2269 & 0.0876 & -0.3987 & -0.0552 & 0.00962 & Statemedia \\
Coverage of Trump asylum for white South Africans with critical response from the South African government   & -0.2078 & 0.0918 & -0.3878 & -0.0278 & 0.02368 & Statemedia \\
China's MOFA accuses the US of espionage against China. & -0.2057 & 0.0703 & -0.3436 & -0.0679 & 0.00345 & Statemedia \\
China's MOFA says that they are opposed Trump's tariffs but support discussion.  & -0.1983 & 0.0774 & -0.3501 & -0.0466 & 0.01042 & Statemedia \\
A Chinese drone using a flamethrower to burn objects caught in high-voltage lines. & -0.1939 & 0.0776 & -0.3461 & -0.0417 & 0.01253 & Statemedia \\
Coverage of Putin at the Moscow Forum, exchanging words in German with a German CEO. & -0.1930 & 0.0753 & -0.3405 & -0.0455 & 0.01032 & Statemedia \\
China's Ministry of National Defense critical of the Japan-US alliance and nuclear threats. & -0.1870 & 0.0816 & -0.3470 & -0.0270 & 0.02201 & Statemedia \\
Coverage of Chinese media firm launch of a media platform to aid expansion of Chinese business and media abroad  & -0.1855 & 0.0920 & -0.3657 & -0.0052 & 0.04371 & Statemedia \\
\bottomrule
\end{tabular}
\end{adjustbox}
\end{table}

%% file: tables/primary_table.tex
\begin{table}[H]
\centering
\resizebox{\ifdim\width>\linewidth 1\linewidth\else\width\fi}{!}{
\begin{talltblr}[         %% tabularray outer open
caption={Intent-to-treat effect of influencer and state media videos on primary outcomes},
note{}={+ p \num{< 0.1}, * p \num{< 0.05}, ** p \num{< 0.01}, *** p \num{< 0.001}},
note{ }={Notes: HC2 robust standard errors in parentheses. Includes covariate adjustment.},
]                     %% tabularray outer close
{                     %% tabularray inner open
colspec={Q[]Q[]Q[]Q[]Q[]Q[]Q[]Q[]Q[]},
column{2-9}={}{halign=c,},
column{1}={}{halign=l,},
hline{8}={1-9}{solid, black, 0.05em},
}                     %% tabularray inner close
\toprule
& Favorability (Forced) & Favorability (Free) & Culture (Forced) & Culture (Free) & Economy (Forced) & Economy (Free) & Politics (Forced) & Politics (Free) \\ \midrule %% TinyTableHeader
Constant & \num{3.20}*** & \num{2.94}*** & \num{0.55}*** & \num{0.59}*** & \num{0.36}** & \num{0.26}*** & \num{0.32}* & \num{0.22}** \\
& (\num{0.13}) & (\num{0.07}) & (\num{0.14}) & (\num{0.07}) & (\num{0.14}) & (\num{0.08}) & (\num{0.14}) & (\num{0.07}) \\
State media & \num{-0.09}+ & \num{-0.09}*** & \num{0.05} & \num{-0.03} & \num{0.07} & \num{0.00} & \num{0.10}+ & \num{0.00} \\
& (\num{0.05}) & (\num{0.03}) & (\num{0.06}) & (\num{0.03}) & (\num{0.05}) & (\num{0.03}) & (\num{0.05}) & (\num{0.03}) \\
Influencers & \num{0.17}*** & \num{0.21}*** & \num{0.24}*** & \num{0.13}*** & \num{0.27}*** & \num{0.23}*** & \num{0.19}*** & \num{0.15}*** \\
& (\num{0.05}) & (\num{0.03}) & (\num{0.05}) & (\num{0.03}) & (\num{0.05}) & (\num{0.03}) & (\num{0.05}) & (\num{0.03}) \\
Num.Obs. & \num{1741} & \num{6698} & \num{1731} & \num{6646} & \num{1737} & \num{6666} & \num{1732} & \num{6655} \\
\bottomrule
\end{talltblr}
}
\end{table}

%% file: tables/secondary_table_forced.tex
\begin{table}[H]
\centering
\resizebox{\ifdim\width>\linewidth 1\linewidth\else\width\fi}{!}{
\begin{talltblr}[         %% tabularray outer open
caption={Intent-to-treat effect of influencer and state media videos on secondary outcomes},
note{}={+ p \num{< 0.1}, * p \num{< 0.05}, ** p \num{< 0.01}, *** p \num{< 0.001}},
note{ }={Notes: HC2 robust standard errors in parentheses. Includes covariate adjustment.},
]                     %% tabularray outer close
{                     %% tabularray inner open
colspec={Q[]Q[]Q[]Q[]Q[]Q[]Q[]Q[]Q[]Q[]Q[]Q[]Q[]},
column{2-13}={}{halign=c,},
column{1}={}{halign=l,},
hline{8}={1-13}{solid, black, 0.05em},
}                     %% tabularray inner close
\toprule
& Admire the Chinese economy & The Chinese economy is advanced & China is a tech leader & Admire the Chinese political system & The Chinese political system is legitimate & The Chinese political system is responsive & Admire Chinese culture & Chinese culture is a positive influence & Want to learn more about Chinese culture & US-China trade cooperation & US-China security cooperation & China is US enemy \\ \midrule %% TinyTableHeader
Constant & \num{4.66}*** & \num{5.44}*** & \num{5.58}*** & \num{3.69}*** & \num{4.65}*** & \num{4.04}*** & \num{5.43}*** & \num{5.45}*** & \num{5.74}*** & \num{5.57}*** & \num{5.12}*** & \num{2.42}*** \\
& (\num{0.24}) & (\num{0.22}) & (\num{0.20}) & (\num{0.25}) & (\num{0.25}) & (\num{0.25}) & (\num{0.23}) & (\num{0.22}) & (\num{0.25}) & (\num{0.22}) & (\num{0.25}) & (\num{0.24}) \\
State media & \num{0.02} & \num{0.11} & \num{0.13} & \num{0.22}* & \num{0.05} & \num{0.24}* & \num{0.03} & \num{0.09} & \num{0.08} & \num{0.22}* & \num{0.21}* & \num{-0.07} \\
& (\num{0.09}) & (\num{0.09}) & (\num{0.08}) & (\num{0.10}) & (\num{0.09}) & (\num{0.10}) & (\num{0.09}) & (\num{0.09}) & (\num{0.10}) & (\num{0.09}) & (\num{0.10}) & (\num{0.09}) \\
Influencers & \num{0.53}*** & \num{0.29}*** & \num{0.29}*** & \num{0.34}*** & \num{0.19}* & \num{0.40}*** & \num{0.33}*** & \num{0.34}*** & \num{0.33}*** & \num{0.27}** & \num{0.26}** & \num{-0.26}** \\
& (\num{0.09}) & (\num{0.08}) & (\num{0.08}) & (\num{0.10}) & (\num{0.09}) & (\num{0.10}) & (\num{0.08}) & (\num{0.08}) & (\num{0.09}) & (\num{0.08}) & (\num{0.09}) & (\num{0.09}) \\
Num.Obs. & \num{1737} & \num{1737} & \num{1737} & \num{1732} & \num{1732} & \num{1732} & \num{1731} & \num{1731} & \num{1731} & \num{1730} & \num{1730} & \num{1730} \\
\bottomrule
\end{talltblr}
}
\end{table}

%% file: tables/secondary_table_free.tex
\begin{table}[H]
\centering
\resizebox{\ifdim\width>\linewidth 1\linewidth\else\width\fi}{!}{
\begin{talltblr}[         %% tabularray outer open
caption={Intent-to-treat effect of influencer and state media videos on secondary outcomes},
note{}={+ p \num{< 0.1}, * p \num{< 0.05}, ** p \num{< 0.01}, *** p \num{< 0.001}},
note{ }={Notes: HC2 robust standard errors in parentheses. Includes covariate adjustment.},
]                     %% tabularray outer close
{                     %% tabularray inner open
colspec={Q[]Q[]Q[]Q[]Q[]Q[]Q[]Q[]Q[]Q[]Q[]Q[]Q[]},
column{2-13}={}{halign=c,},
column{1}={}{halign=l,},
hline{8}={1-13}{solid, black, 0.05em},
}                     %% tabularray inner close
\toprule
& Admire the Chinese economy & The Chinese economy is advanced & China is a tech leader & Admire the Chinese political system & The Chinese political system is legitimate & The Chinese political system is responsive & Admire Chinese culture & Chinese culture is a positive influence & Want to learn more about Chinese culture & US-China trade cooperation & US-China security cooperation & China is US enemy \\ \midrule %% TinyTableHeader
Constant & \num{4.45}*** & \num{5.29}*** & \num{5.55}*** & \num{3.76}*** & \num{4.18}*** & \num{4.01}*** & \num{5.51}*** & \num{5.36}*** & \num{5.92}*** & \num{5.67}*** & \num{5.12}*** & \num{2.43}*** \\
& (\num{0.13}) & (\num{0.12}) & (\num{0.11}) & (\num{0.13}) & (\num{0.13}) & (\num{0.13}) & (\num{0.12}) & (\num{0.11}) & (\num{0.12}) & (\num{0.12}) & (\num{0.12}) & (\num{0.13}) \\
State media & \num{-0.09}+ & \num{0.03} & \num{0.05} & \num{-0.05} & \num{-0.00} & \num{0.06} & \num{-0.04} & \num{-0.02} & \num{-0.06} & \num{0.06} & \num{0.06} & \num{0.04} \\
& (\num{0.05}) & (\num{0.04}) & (\num{0.04}) & (\num{0.05}) & (\num{0.05}) & (\num{0.05}) & (\num{0.05}) & (\num{0.04}) & (\num{0.05}) & (\num{0.04}) & (\num{0.05}) & (\num{0.05}) \\
Influencers & \num{0.37}*** & \num{0.30}*** & \num{0.24}*** & \num{0.21}*** & \num{0.22}*** & \num{0.28}*** & \num{0.20}*** & \num{0.21}*** & \num{0.15}*** & \num{0.11}* & \num{0.09}+ & \num{-0.08}+ \\
& (\num{0.05}) & (\num{0.04}) & (\num{0.04}) & (\num{0.05}) & (\num{0.05}) & (\num{0.05}) & (\num{0.04}) & (\num{0.04}) & (\num{0.05}) & (\num{0.04}) & (\num{0.05}) & (\num{0.05}) \\
Num.Obs. & \num{6666} & \num{6666} & \num{6665} & \num{6655} & \num{6655} & \num{6654} & \num{6646} & \num{6645} & \num{6645} & \num{6640} & \num{6640} & \num{6641} \\
\bottomrule
\end{talltblr}
}
\end{table}

%% file: tables/ri_pvalues.tex
\begin{table}[H]
\centering
\resizebox{\ifdim\width>\linewidth 0.9\linewidth\else\width\fi}{!}{
\begin{talltblr}[         %% tabularray outer open
caption={Randomization inference p-values \label{tab: ri}},
note{}={Note: Calculated using 10,000 simulations},
]                     %% tabularray outer close
{                     %% tabularray inner open
colspec={Q[]Q[]Q[]Q[]Q[]},
}                     %% tabularray inner close
\toprule
Outcome & Free choice: state media & Free choice: influencers & Forced: state media & Forced: influencers \\ \midrule %% TinyTableHeader
Culture & 0.2828 & 0.0000 & 0.2491 & 0.0000 \\
Economy & 0.9746 & 0.0000 & 0.1278 & 0.0000 \\
Favorability & 0.0006 & 0.0000 & 0.1962 & 0.0001 \\
Petition interest & 0.3883 & 0.5671 & 0.8246 & 0.9699 \\
Petition signing & 0.3234 & 0.6769 & 0.3931 & 0.1595 \\
Politics & 0.9629 & 0.0000 & 0.0351 & 0.0002 \\
China is US enemy & 0.3334 & 0.1203 & 0.3674 & 0.0025 \\
US-China security cooperation & 0.2696 & 0.0793 & 0.0132 & 0.0045 \\
US-China trade cooperation & 0.2093 & 0.0223 & 0.0062 & 0.0013 \\
\bottomrule
\end{talltblr}
}
\end{table}

%% file: tables/mcc.tex
\begin{table}
\centering
\resizebox{\ifdim\width>\linewidth 0.9\linewidth\else\width\fi}{!}{
\begin{talltblr}[         %% tabularray outer open
caption={Estimates and multiple comparisons corrections, all secondary and exploratory outcomes \label{tab: mcc}},
note{}={p values less than 0.05 shaded green, p values greater than 0.05 and less than 0.1 shaded orange, and p-values greater than 0.1 shaded red.},
]                     %% tabularray outer close
{                     %% tabularray inner open
colspec={Q[]Q[]Q[]Q[]Q[]Q[]Q[]Q[]Q[]Q[]Q[]},
cell{16}{6}={}{bg=cFFA500,},
cell{16}{7}={}{bg=cFFA500,},
cell{16}{8}={}{bg=cFFA500,},
cell{16}{9}={}{bg=cFFA500,},
cell{17-18}{11}={}{bg=cFFA500,},
cell{17-53}{6}={}{bg=cFF0000,},
cell{17-53}{7}={}{bg=cFF0000,},
cell{17-53}{8}={}{bg=cFF0000,},
cell{17-53}{9}={}{bg=cFF0000,},
cell{19-53}{11}={}{bg=cFF0000,},
cell{2-15}{6}={}{bg=c00FF7F,},
cell{2-15}{7}={}{bg=c00FF7F,},
cell{2-15}{8}={}{bg=c00FF7F,},
cell{2-15}{9}={}{bg=c00FF7F,},
cell{2-16}{11}={}{bg=c00FF7F,},
cell{2-23}{10}={}{bg=c00FF7F,},
cell{2-25}{5}={}{bg=c00FF7F,},
cell{24-25}{10}={}{bg=cFFA500,},
cell{26-29}{5}={}{bg=cFFA500,},
cell{26-53}{10}={}{bg=cFF0000,},
cell{30-53}{5}={}{bg=cFF0000,},
}                     %% tabularray inner close
\tinytableDefineColor{cFF0000}{HTML}{FF0000}
\tinytableDefineColor{cFFA500}{HTML}{FFA500}
\tinytableDefineColor{c00FF7F}{HTML}{00FF7F}
\toprule
Forced or free & Content type & Outcome & Estimate & Unadjusted & Bonferroni & Holm & Hochberg & Hommel & Benjamini Hochberg & Benjamini Yekutieli \\ \midrule %% TinyTableHeader
Free & State media & Inspired & -0.37 & 0.000 & 0.000 & 0.000 & 0.000 & 0.000 & 0.000 & 0.000 \\
Free & State media & Excited & -0.35 & 0.000 & 0.000 & 0.000 & 0.000 & 0.000 & 0.000 & 0.000 \\
Free & State media & Upset & 0.26 & 0.000 & 0.000 & 0.000 & 0.000 & 0.000 & 0.000 & 0.000 \\
Free & Influencers & Economy & 0.23 & 0.000 & 0.000 & 0.000 & 0.000 & 0.000 & 0.000 & 0.000 \\
Forced & State media & Upset & 0.44 & 0.000 & 0.000 & 0.000 & 0.000 & 0.000 & 0.000 & 0.000 \\
Forced & State media & Inspired & -0.44 & 0.000 & 0.000 & 0.000 & 0.000 & 0.000 & 0.000 & 0.000 \\
Forced & State media & Excited & -0.41 & 0.000 & 0.000 & 0.000 & 0.000 & 0.000 & 0.000 & 0.000 \\
Free & Influencers & Politics & 0.15 & 0.000 & 0.000 & 0.000 & 0.000 & 0.000 & 0.000 & 0.000 \\
Forced & Influencers & Economy & 0.27 & 0.000 & 0.000 & 0.000 & 0.000 & 0.000 & 0.000 & 0.000 \\
Free & Influencers & Culture & 0.13 & 0.000 & 0.000 & 0.000 & 0.000 & 0.000 & 0.000 & 0.000 \\
Forced & Influencers & Culture & 0.24 & 0.000 & 0.001 & 0.000 & 0.000 & 0.000 & 0.000 & 0.000 \\
Forced & Influencers & Foreign policy & 0.21 & 0.000 & 0.010 & 0.008 & 0.008 & 0.007 & 0.001 & 0.004 \\
Forced & Influencers & Politics & 0.19 & 0.000 & 0.016 & 0.012 & 0.012 & 0.012 & 0.001 & 0.005 \\
Free & State media & Interested & -0.13 & 0.000 & 0.025 & 0.018 & 0.018 & 0.018 & 0.002 & 0.008 \\
Forced & Influencers & US-China trade cooperation & 0.27 & 0.002 & 0.084 & 0.062 & 0.062 & 0.062 & 0.006 & 0.025 \\
Forced & Influencers & China is US enemy & -0.26 & 0.004 & 0.222 & 0.158 & 0.158 & 0.142 & 0.014 & 0.063 \\
Forced & Influencers & US-China security cooperation & 0.26 & 0.005 & 0.286 & 0.198 & 0.198 & 0.176 & 0.017 & 0.076 \\
Free & Influencers & Foreign policy & 0.08 & 0.010 & 0.509 & 0.342 & 0.342 & 0.293 & 0.028 & 0.126 \\
Forced & State media & Foreign policy & 0.15 & 0.010 & 0.528 & 0.345 & 0.345 & 0.304 & 0.028 & 0.126 \\
Forced & State media & US-China trade cooperation & 0.22 & 0.011 & 0.589 & 0.374 & 0.374 & 0.340 & 0.029 & 0.134 \\
Free & Influencers & US-China trade cooperation & 0.11 & 0.013 & 0.666 & 0.410 & 0.410 & 0.384 & 0.032 & 0.144 \\
Free & Influencers & Inspired & -0.09 & 0.017 & 0.870 & 0.519 & 0.519 & 0.485 & 0.040 & 0.180 \\
Forced & Influencers & Interested & 0.16 & 0.025 & 1.000 & 0.746 & 0.746 & 0.686 & 0.056 & 0.255 \\
Forced & State media & US-China security cooperation & 0.21 & 0.027 & 1.000 & 0.772 & 0.772 & 0.719 & 0.058 & 0.262 \\
Forced & State media & Politics & 0.10 & 0.059 & 1.000 & 1.000 & 0.982 & 0.982 & 0.123 & 0.557 \\
Free & Influencers & US-China security cooperation & 0.09 & 0.068 & 1.000 & 1.000 & 0.982 & 0.982 & 0.137 & 0.621 \\
Free & Influencers & Excited & -0.06 & 0.097 & 1.000 & 1.000 & 0.982 & 0.982 & 0.182 & 0.827 \\
Free & Influencers & China is US enemy & -0.08 & 0.098 & 1.000 & 1.000 & 0.982 & 0.982 & 0.182 & 0.827 \\
Free & State media & US-China trade cooperation & 0.06 & 0.149 & 1.000 & 1.000 & 0.982 & 0.982 & 0.265 & 1.000 \\
Forced & Influencers & Petition signing & 0.02 & 0.153 & 1.000 & 1.000 & 0.982 & 0.982 & 0.265 & 1.000 \\
Forced & State media & Economy & 0.07 & 0.213 & 1.000 & 1.000 & 0.982 & 0.982 & 0.357 & 1.000 \\
Free & State media & US-China security cooperation & 0.06 & 0.228 & 1.000 & 1.000 & 0.982 & 0.982 & 0.370 & 1.000 \\
Free & State media & Foreign policy & 0.03 & 0.250 & 1.000 & 1.000 & 0.982 & 0.982 & 0.394 & 1.000 \\
Free & State media & Petition signing & 0.01 & 0.295 & 1.000 & 1.000 & 0.982 & 0.982 & 0.451 & 1.000 \\
Free & State media & Culture & -0.03 & 0.322 & 1.000 & 1.000 & 0.982 & 0.982 & 0.478 & 1.000 \\
Forced & Influencers & Upset & 0.05 & 0.365 & 1.000 & 1.000 & 0.982 & 0.982 & 0.527 & 1.000 \\
Free & State media & China is US enemy & 0.04 & 0.390 & 1.000 & 1.000 & 0.982 & 0.982 & 0.546 & 1.000 \\
Forced & State media & Petition signing & -0.01 & 0.399 & 1.000 & 1.000 & 0.982 & 0.982 & 0.546 & 1.000 \\
Free & State media & Petition interest & -0.01 & 0.418 & 1.000 & 1.000 & 0.982 & 0.982 & 0.557 & 1.000 \\
Forced & State media & Culture & 0.05 & 0.432 & 1.000 & 1.000 & 0.982 & 0.982 & 0.561 & 1.000 \\
Forced & State media & China is US enemy & -0.07 & 0.447 & 1.000 & 1.000 & 0.982 & 0.982 & 0.567 & 1.000 \\
Free & Influencers & Petition interest & 0.01 & 0.513 & 1.000 & 1.000 & 0.982 & 0.982 & 0.635 & 1.000 \\
Free & Influencers & Upset & 0.02 & 0.527 & 1.000 & 1.000 & 0.982 & 0.982 & 0.637 & 1.000 \\
Free & Influencers & Interested & 0.02 & 0.539 & 1.000 & 1.000 & 0.982 & 0.982 & 0.637 & 1.000 \\
Forced & Influencers & Inspired & -0.04 & 0.604 & 1.000 & 1.000 & 0.982 & 0.982 & 0.698 & 1.000 \\
Free & Influencers & Petition signing & 0.00 & 0.756 & 1.000 & 1.000 & 0.982 & 0.982 & 0.855 & 1.000 \\
Forced & State media & Petition interest & -0.01 & 0.791 & 1.000 & 1.000 & 0.982 & 0.982 & 0.876 & 1.000 \\
Forced & State media & Interested & 0.02 & 0.831 & 1.000 & 1.000 & 0.982 & 0.982 & 0.901 & 1.000 \\
Forced & Influencers & Excited & -0.01 & 0.890 & 1.000 & 1.000 & 0.982 & 0.982 & 0.945 & 1.000 \\
Free & State media & Economy & 0.00 & 0.925 & 1.000 & 1.000 & 0.982 & 0.982 & 0.962 & 1.000 \\
Free & State media & Politics & 0.00 & 0.954 & 1.000 & 1.000 & 0.982 & 0.982 & 0.972 & 1.000 \\
Forced & Influencers & Petition interest & 0.00 & 0.982 & 1.000 & 1.000 & 0.982 & 0.982 & 0.982 & 1.000 \\
\bottomrule
\end{talltblr}
}
\end{table}